\documentclass[aps,prb,twocolumn,superscriptaddress,nolinenumbers,nobibnotes,nodoi,noeprint]{revtex4-2}

\usepackage{amsmath,amssymb,physics,bm,siunitx}
\usepackage{xcolor,graphicx}
\usepackage{textcomp}
\usepackage{ulem}
\usepackage[papersize={8.5in,11in}]{geometry}
\usepackage[colorlinks=true]{hyperref}
\hypersetup{
    bookmarks=true,         % show bookmarks bar?
    unicode=false,          % non-Latin characters 
    pdftoolbar=true,        % show Acrobat
    pdfmenubar=true,        % show Acrobat 
    pdffitwindow=false,     % window fit to page when opened
    pdfstartview={FitH},    % fits the width of the page to the window
    pdfkeywords={keyword1} {key2} {key3}, % list of keywords
    pdfnewwindow=true,      % links in new window
    colorlinks=true,       % false: boxed links; true: colored links
    linkcolor=magenta, %red,          % color of internal links (change box color with linkbordercolor)
    citecolor=blue,        % color of links to bibliography
    filecolor=magenta,      % color of file links
    urlcolor=blue           % color of external links
} 

\definecolor{modernorange}{HTML}{FF6B35}
\definecolor{darkblue}{HTML}{0066CC}
\definecolor{tealmodern}{HTML}{006D77}
\definecolor{slate}{HTML}{37474F}  
\definecolor{plum}{HTML}{6A4C93}

\AtBeginDocument{\RenewCommandCopy\qty\SI}
\begin{document}

\title{Anyonic time exchange in hierarchical Hall states: emergent Luttinger scaling and cross-correlations at finite temperature}
\author{Aleksander Latyshev}
\author{In\`es Safi}
\affiliation{Laboratoire de Physique des Solides (UMR 5802), CNRS-Universit\'e Paris-Saclay,
B\^atiment 510, 91405 Orsay, France}

\begin{abstract}
We address hierarchical fractional quantum Hall edges coupled by a spatially local quantum point contact within the Unified Nonequilibrium Perturbative (UNEP) framework. Without assuming a Tomonaga-Luttinger liquid (TLL) form for the tunneling correlator, we impose a local anyonic time-exchange (ATE) constraint with phase $\bar{\theta}$ that results in a nonequilibrium fluctuation-dissipation integral equation connecting the DC tunneling current and noise. For thermal initial states, detailed balance and the weak-tunneling Poisson relation make this integral equation a closed one for the current. Using the Wiener-Hopf method, we find a unique physical solution of the local TLL form with a scaling dimension $\delta$ determined by $\bar{\theta}=\pi\delta\pmod{\pi}$ which emerges from this solution rather than being imposed through a TLL correlator. By contrast, ATE combined only with inversion symmetry fixes the exchange monodromy but does not determine the complete transport dynamics. We then consider the "anyon collider" geometry beyond Laughlin states, where detailed balance is broken and the noise becomes super-Poissonian. The relation between the nonequilibrium and equilibrium correlators determines the finite-temperature current and noise. We further derive the measurable output cross-correlation and the balanced generalized Fano factor, which depend both on the scaling dimension $\delta$ and on the statistical phase $\theta$,  identical to $\bar{\theta}$ for co-propagating modes with short-range interactions. 
We identify a temperature-dependent current scale separating
 the zero-temperature asymptotic regime from the low-current regime at which the balanced Fano factor saturates to a function of $\theta$ and $\delta$
distinct from its zero-temperature value. The latter regime turns out to be particularly relevant to experiments on "anyon colliders." These complementary
regimes provide additional constraints on the statistical phase
and scaling dimension, subject to the phase ambiguity
$\theta\mapsto\theta+\pi$ and the symmetry
$\delta\mapsto1-\delta$ at fixed $\theta$. For suitable independent values of the statistical phase and scaling dimension, the thermal crossover can reverse the sign of the output cross-correlation without changing quasiparticle statistics.
\end{abstract}

%\patiocs{}

\maketitle

\section{Introduction} Anyons, quasiparticles exhibiting exchange phases intermediate between bosons and fermions, are expected to emerge in the fractional quantum Hall effect (FQHE)~\cite{laughlin_fractional_charge_PRL_83,fractional_statistics_anyons_leinaas_myrheim_1977}. Their fractional statistics manifest through quantum correlations and transport signatures, the probing of which remains a central challenge. Two main lines of investigation involving two or more Quantum Point Contacts (QPCs) have focused on braiding either in the spatial or temporal domain, often assumed to probe the same statistical phase $\theta$ \cite{fractional_statistics_theory_Sim_Nat_comm_2016,fractional_statistics_theory_2016,oreg_PRL_FQHE_statistics_2023,fractional_statistics_rosenow_halperin_proceedings_2025}, though it has been shown that the temporal braiding phase is generally less protected than the spatial one  and can split into nonuniversal components \cite{ines_statistics_2025,fractional_statistics_christian_splatt_FQHE_2026} . However, interpreting these signatures in hierarchical Hall states remains challenging, particularly because the quasiparticle statistical phase $\theta$ is not generally fixed by the filling factor, unlike in Laughlin states.

The first line is realized in interferometric setups, where interference between paths enclosing a magnetic flux reveals the statistical phase~\cite{manfra_FQHE_statistics_Naturephys_2020,manfra_2_5_braiding_FPI_FQHE_PRX_2023,fractional_statistics_anyons_FQHE_heiblum_MZI_2_5_Nature_physics_2023,sukho_mac_zhender_PRB_2009,fractional_statistics_noise_FPI_FQHE_rosenow_PRB_2012}. Such schemes are, however, notoriously sensitive to Coulomb interactions~\cite{manfra_braiding_anyons_2022} and device-specific parameters, which complicate their interpretation and limit their universality~\cite{fractional_statistics_comment_Read_Nat_Phys_2023,fractional_statistics_comment_Kivelson_Journal_club_2023}.  

The second line is based on current cross-correlations such as Hanbury-Brown and Twiss setups~\cite{hbt,kim_hbt,*vish_hall_hbt_2003,gefen_HBT_FQHE_PRL_2012,*giuliano_HBT_FQHE_2016,giuliano_HBT_FQHE_PRB_2024} or the so-called "anyon collider" \footnote{For convenience, we use this terminology even though the diluted beams of anyons do not literally collide at the central QPC.} geometry~\cite{fractional_statistics_theory_2016,fractional_statistics_rosenow_halperin_proceedings_2025}  achieved in pioneering experiments~\cite{fractional_statistics_gwendal_science_2020,fractional_statistics_heiblum_sim_nature_2023} interpreted in terms of anyon braiding in the time domain~\cite{fractional_statistics_theory_Sim_Nat_comm_2016,fractional_statistics_Sim_PRL_negative_shot_noise_PRL_2019,Sim_non_abelian_NATcomm_2022,mora2022anyonicexchangebeamsplitter,fractional_satistics_zhang_gefen_PRL_2025,fractional_statistics_zhang_Gefen_2025_anyon_collider}.  

Yet the interpretation of these experiments largely relies on the Tomonaga-Luttinger liquid (TLL) framework with a nonuniversal scaling
dimension $\delta$ \footnote{The scaling dimension $\delta$ is defined from the Green’s function of the quasiparticle field $\Psi$ as $\langle \Psi^{\dagger}(x,t)\Psi(0,0)\rangle \simeq  (x-vt )^{-\delta}$, where $v$ is the plasmon velocity. In the zero-temperature limit, the DC current, obtained from the Fourier transform of the correlation functions of a spatially local tunneling operator $A$, scales as $\omega_{\scriptscriptstyle\mathrm dc} ^{2\delta - 1}$, thus increasing as the scaling dimension $\delta$ decreases. 
}, without providing a direct verification of the TLL prediction for DC current. In addition, cross-correlations computed at zero temperatures cannot disentangle the time-domain braiding phase ${\theta}$  from $\delta$.  Experimental results in Jain states ($\nu=2/5$) even show quantitative discrepancies with theoretical expectations~\cite{pierre_anyons_PRX_2023,fractional_statistics_gwendal_PRX_2023,fractional_statistics_width_anyons_rosenow_PRL_2024,martin_anyons_finite_width_FQHE_PRL_2024}.

This motivates the pursuit of more direct and unambiguous probes of anyonic statistics~\cite{fractional_statistics_Sim_PRL_negative_shot_noise_PRL_2019,Sim_non_abelian_NATcomm_2022,vish_FQHE_single_edge_PRB_2023,fractional_statistics_heiblum_sim_nature_2023,oreg_PRL_FQHE_statistics_2023}, which have so far been mostly explored within the TLL framework, and, in some cases, simplified to a single quantum point contact (QPC) configuration
 ~\cite{vish_FQHE_single_edge_PRB_2023}. In particular, based on \cite{martin_fractional_statistics_dip_HOM_PRL_2023} which analyzes the width of the Hong-Ou-Mandel  dip for time-controlled anyons generated by voltage pulses (as originally suggested in ~\cite{ines_ann}), the experimental work in ~\cite{feve_anyons_martin_Science_2025} reports a disentangled determination of the scaling parameter $\delta$ from the braiding phase in the time domain expected to be universal, unlike $\delta$. However, the analysis relies on an assumed TLL form of the tunneling
correlator, whose scaling was not independently verified in that
experiment. The theoretical interpretation has subsequently been
reexamined in
Refs.~\cite{matteo_HOM_FQHE_anyons_PRB_2025,ines_alex_2025}.

Recent advances formulated within a unified nonequilibrium perturbative (UNEP) framework~\cite{ines_eugene,*ines_cond_mat,ines_degiovanni_2016,*ines_PRB_R_noise_2020,ines_photo_noise_PRB_2022} have suggested three alternative protocols for accessing the anyonic phase valid for a nonequilibrium distribution, for instance, due to biased temperatures, but with different minimal conditions. First, Ref.~\cite{ines_PRB_R_noise_2020} is based on the determination of the effective voltage drop at the central QPC in an "anyon collider'' that does not require the TLL model \footnote{A similar idea underlies Ref. \cite{fractional_statistics_zhang_Gefen_2025_anyon_collider},  however, it is based on the TLL model)}.
Secondly, Ref.~\cite{ines_lucas_Seddik_PRB_2026} is based on an injected time-resolved anyon that induces dynamic oscillations of the transmitted charge through a single QPC.
Thirdly, Ref. ~\cite{ines_statistics_2025}, which is the starting point of the present paper, considers states beyond Laughlin one described by low-energy effective theories \cite{wen_FQHE_effective_theories_PRB_1990} within both single-QPC and "anyon collider" setups, without a specific injection prescription or the preparation of a particular nonequilibrium distribution. These protocols aim to test an anyonic time-exchange (ATE) link between tunneling operators involving the anyonic phase $\theta$. While the ATE link is necessary for anyon braiding in the time domain and likewise depends on the locality of the QPC, it has been shown to be more robust against short-range  interactions between co-propagating modes (on each edge separately), for which one has the identity $\theta = \pi \delta$,  $\delta$ being the scaling dimension. However, this identification breaks down for counter-propagating modes or with interactions between the upper and lower edges, for which the ATE link holds with $\pi\delta$ replacing $\theta$, so that these protocols remain promising for the determination of $\delta$. 
When considering the exchange of dynamical anyon fields at different positions, interactions between $N$ edge modes become crucial: they fractionalize~\cite{ines_ann} an injected anyon into $N$ components, each carrying a nonuniversal effective phase, whose sum remains equal to the statistical phase $\theta$\footnote{See also the more recent work in Ref.~\cite{fractional_statistics_christian_splatt_FQHE_2026}.}. \cite{ines_statistics_2025} Consequently, time-domain braiding, which requires co-propagating modes, generally involves the successive braiding of these fractionalized components with the anyon--anti-anyon excitation at the QPC, rather than a single anyon braiding with a phase $\theta$. The standard time-domain braiding scenario is recovered only for equal mode velocities  \cite{oreg_PRL_FQHE_statistics_2023} or sufficiently local injection near the QPC, which suppresses the spreading of the fractionalized components.
%An injected anyon in edge states with short-range inter-mode interactions is farctionnali of into non universal components with nonuniversal statistical phases fractional charge }
%Importantly, the identity  $\bar{\theta}=\theta=\pi \delta (mod \; \pi)$ has been established in the presence of edge interactions on each side of the Hall  bar and for a spatially local QPC \cite{ines_statistics_2025}. Thus $\delta$ turns out to be more robust with respect to such edge interactions than currently believed, due to the locality of the QPC }. 
A second key insight of Ref.~\cite{ines_statistics_2025} is that the ATE link, being robust against fractionalization, yields nonequilibrium fluctuation--dissipation relations applicable to both single- and multi-QPC geometries, including configurations with temperature biases. They relate the DC shot noise, either through an integral equation, to the DC current average, or to the admittance phase. These relations provide novel, robust methods to test the validity of the ATE link, as well as to isolate the role of  ${\theta}$  as an exchange phase from its possible role as a scaling dimension.
%This conceptual advance indicates %that the anyonic braiding phase in the time domain can, in principle, %be inferred using a single QPC at \IS{one or different finite temperatures}, without requiring "anyon %collider" geometries \IS{restricted to zero temperature}, although the same relations remain valid there as %well.

The paper is organized as follows. Section~I summarizes our main results. In Sec.~II, we review the UNEP framework and the integral relations between DC current and noise imposed by the ATE constraint. In Sec.~III, we consider thermal initial states, impose detailed balance, and solve the resulting Poissonian-noise integral equation using the Wiener--Hopf (W--H) method, establishing the uniqueness and TLL-type scaling of the physical solution. In Sec.~IV, we apply the framework to the ``anyon collider'' geometry, deriving the finite-temperature tunneling current and noise, output cross-correlations, and generalized Fano factor for balanced injection. We analyze its low-temperature and low-injected-current limits and propose new protocols to disentangle $\theta$ and $\delta$, subject to the ambiguity $\theta\mod\pi$ and, when $\theta$ and $\delta$ are independent, the additional ambiguity $\delta\leftrightarrow 1-\delta$. Finally, Sec.~V summarizes our conclusions and discusses their implications for time-domain braiding.
\section{Overview}
In the present paper, we search for alternative candidates to low-energy effective models that satisfy the ATE constraint  involving a general ATE phase $\bar{\theta}$, without specifying the underlying Hamiltonian within the UNEP framework. When $\bar{\theta}=\theta$, since the ATE constraint is a prerequisite for ensuring anyon braiding in the time domain under its associated conditions and involves the same phase, our approach importantly delimits the models for which the latter can indeed be ensured and answers, more generally, the question: can one demonstrate ATE without evidence for TLL behavior? For that, we solve the integral equation relating the DC noise to the DC current. We first focus on a single QPC initialized in thermal equilibrium, so that the nonequilibrium DC noise is Poissonian and the  integral equation concerns a unique observable: the DC current. We use the Wiener-Hopf (W-H) technique \cite{noble1958wiener,gogolin2014lectures}, uncovering deep analogies between the time-domain braiding phase and the analytic properties of the response function. Building upon this framework, we demonstrate that there is a unique solution given by the local TLL-type DC current with a scaling dimension $\delta=\bar{\theta}/\pi-m$ where the integer $m$ selects the principal branch $0<\delta<1$. This signature is not an indication of a free chiral TLL model for the edge channels, between which one can still have Coulomb interactions; the robustness of the TLL behavior relies on the ATE structure imposed locally on the QPC. Building on these results, we turn to the dilute-source "anyon collider" prescription, where diluted quasiparticle beams emitted by two independent QPC sources are partitioned at a central analyzer QPC. The injected currents break detailed balance, introducing, in addition to their difference $I_{-}$, a second nonequilibrium scale given by their sum $I_{+}$.  This gives rise to super-Poissonian tunneling noise attributed more generally, within UNEP theory, to nonequilibrium distributions~\cite{ines_cond_mat,ines_PRB_R_noise_2020}.  Then, the fixed relation between the non-equilibrium "collider" correlators and the thermal equilibrium ones allows us to determine the finite-temperature tunneling current and noise once the unique TLL solution, obtained for thermal correlators, is taken into account. We further derive the full cross-correlation and the corresponding balanced Fano factor depending on both the statistical phase $\theta$ and the scaling dimension $\delta$, as the analysis of Ref.~\cite{ines_statistics_2025} shows that $I_{\pm}$ are renormalized in a robust fashion by functions of $\theta$ independently of inter-mode interactions.

%We express the balanced Fano factor a}t $I_-=0$, recovering ${\mathcal F}_0$, \textbf{the known zero-temperature function  of $\theta,\delta$,} \textbf{which acquires} nontrivial thermal corrections. \textbf{In particular, the linear slope with respect to $\omega_{\rm th}/I_+$, where} \textbf{$\omega_{\rm th}=k_{\rm B}T/\hbar$, depends only on $\theta$ and not on $\delta$, while the second-order term provides a different function of $\theta,\delta$ } \textbf{from ${\mathcal F}_0$, thus allowing a determination of both unknown quantities. However, we show that the opposite regime of low injected currents (or relatively high temperature) has been the most explored experimentally, as low but finite temperatures are unavoidable, while the  injected current is arbitrarily low}. \textbf{ In this regime}, the balanced Fano factor reaches a constant value $1-\tan\theta\,R_\delta$ \textbf{easier to extract than low-temperature corrections to $\mathcal{F}_0$, though $\delta$-dependence of $R_\delta$ is more complicated than that of $\mathcal{F}_0$} \sout{due to a cancelation between the leading equilibrium part of the tunneling noise and the mixed thermal contribution. Depending on \(\theta\) and \(\delta\), the thermal crossover may also induce a sign reversal of the cross-correlation.}\textbf{\textit{ I don't think this remark will be understood at this stage}}

We derive the balanced Fano factor at $I_-=0$ as a function of $e^*\omega_{th}/I_+$ (with $\omega_{\rm th}=k_{\rm B}T/\hbar$), showing its symmetry with respect to $\delta\leftrightarrow1-\delta$ (whenever $\theta$ is independent of $\delta$), in addition to the standard symmetry $\theta\rightarrow \theta+\pi$. 
We recover $\mathcal{F}_0$, the zero-temperature function
of $\theta$ and $\delta$ in Ref.\cite{fractional_statistics_theory_2016}, and furthermore determine its thermal corrections.
In particular, the leading low temperature correction is given by
$2e^*\omega_{\rm th}/\sin(2\theta)I_+$
thus depends only on $\theta$ and not on $\delta$  whereas the
second-order term provides a function of $\theta$
and $\delta$ distinct from $\mathcal{F}_0$. These results provide
complementary constraints on the two parameters. The complementary low-current regime can be approached by reducing
$I_+$ at a fixed and unavoidable nonzero temperature while
maintaining dilute injection and perturbative analyzer tunneling.
This regime is particularly relevant to experimental works with dilute
anyon injection where the  injected current is made arbitrarily low.
Then, the generalized Fano factor approaches a constant
value $1-\tan\theta\,R_\delta$ which is easier to extract than low-temperature corrections to $\mathcal{F}_0$, though the $\delta$-dependence of $R_\delta$ is more complicated than that of $\mathcal{F}_0$. This limit
provides an additional constraint on $\theta$ and $\delta$ up to the ambiguity $\theta \mod \pi$ and  $\delta\rightarrow1-\delta$.  At fixed nonzero temperature and balanced injection,
the output cross-correlation vanishes as $I_+\to0$,
consistently with the equilibrium fluctuation-dissipation
theorem. The generalized Fano factor nevertheless
approaches a finite limit, since its numerator and
denominator both vanish linearly with $I_+$.

Therefore, the temperature and injected-current dependencies provide complementary information on the statistical phase $\theta$ and the scaling dimension $\delta$. While the magnitude of the thermal crossover depends on both parameters, its current scale is controlled by $\omega_+=2\sin^2\theta\,I_+/e^*$, which depends on $\theta$ independently of $\delta$. Their combined analysis can help, therefore, to disentangle $\theta$ and $\delta$, up to the phase ambiguity and the symmetry $\delta\rightarrow1-\delta$. The possibility for zero crossing  of the Fano factor as a function of temperature or, equivalently, of the injected current, provides a qualitative diagnostic of a separation between $\theta$ and $\pi\delta$ within the present model.

% The ATE fluctuation-dissipation relations provide an amplitude-independent route to $\bar{\theta}$ by combining broadband measurements of the DC current and noise. Whenever $\bar{\theta}$ \textbf{is identical to the statistical phase $\theta$, the latter can then be determined by this procedure.} \sout{In the collider geometry, the \textbf{linear low-temperature correction of the balanced Fano factor depends only on $\theta $, thus providing a method to disentangle $\theta\mod \pi$ %from $\delta$} , while a zero crossing provides a qualitative diagnostic of a separation between $\theta$ and $\pi\delta$ within the %present model.}
%\textbf{ I changed the preceding paragraph, forgetting you have inserted this one that I suppressed; but please free to %maintian it or  fuse ?}}

The ATE fluctuation--dissipation relations provide a separate route
to the phase $\bar{\theta}$ modulo $\pi$ by combining measurements
of the DC tunneling current and noise over a sufficiently broad
frequency range. This determination is independent of the overall
tunneling amplitude. Whenever $\bar{\theta}$ coincides with the
statistical phase $\theta$, the same procedure gives access to
$\theta$ modulo $\pi$.

%\ipl{Here, ``high-temperature behavior'' should be replaced by
%``linear low-temperature correction''. Indeed, the high-temperature
%limit $\mathcal F_\infty=1-\tan\theta\,R_\delta$ depends on both
%$\theta$ and $\delta$. It is the coefficient of the linear
%low-temperature correction in Eq.~(66) that is independent of $\delta$.}

Our results establish, therefore, the hierarchy between exchange and transport: ATE alone fixes the local exchange monodromy, thermal detailed balance selects the physical TLL-type solution, and the nonequilibrium collider prescription introduces additional scales that make the distinct roles of the statistical phase and scaling dimension directly visible in finite-temperature cross-correlations.

\section{Derivation Recap: ATE Integral Equation}
\begin{figure}[t!]
\centering
\includegraphics[width= 0.99 \linewidth]{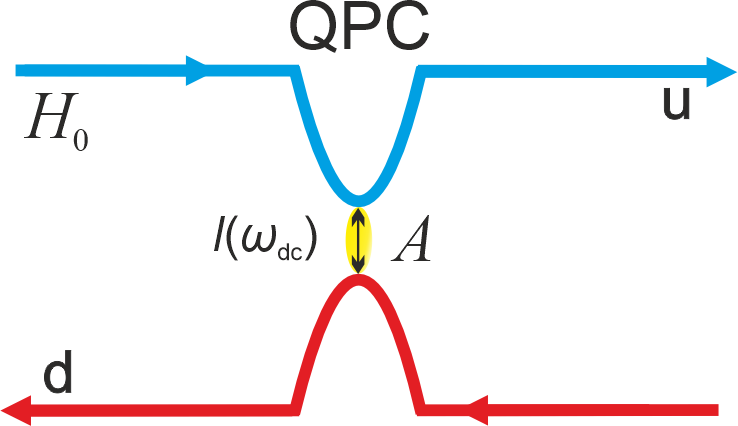}
\caption{A spatially local QPC with a tunneling (or anyon tunneling) operator $A$ between the chiral upper (u) and lower edges (d) that are described by a Hamiltonian $H_0$.}
\label{fig.QPC}
\end{figure}

This section recalls and reformulates some aspects of Ref.~\cite{ines_statistics_2025}. At the first step, we consider the Hamiltonian of the UNEP theory in the stationary regime:
\begin{equation}\label{eq:hamiltonian}
    {H}(t) ={H}_0 + e^{-i\omega_{\scriptscriptstyle\mathrm{dc}} t} A + e^{i\omega_{\scriptscriptstyle\mathrm{dc}} t} A^{\dagger},
\end{equation}
where $\omega_{\scriptscriptstyle\mathrm{dc}}$ can generally be regarded as a free parameter. For generality, we do not rely on any explicit identification of $H_0,A$  nor the initial nonequilibrium and stationary density matrix $\rho_{\mathrm{neq}}$. In addition to the minimal conditions of the UNEP theory recalled in Appendix~\ref{app:conditions}, mainly a weak amplitude of $A$, we add the ATE constraint:
 $\langle A^{\dagger}(t)A(0) \rangle = e^{-2 i \bar{\theta} \, \text{sign}(t)} \langle A(0)A^{\dagger}(t) \rangle$ which reads as
\begin{equation}\label{eq:braiding_X}
X_{\scriptscriptstyle\mathrm{\rightarrow}}(t)=e^{-2 i \bar{\theta} \, \text{sign}(t) }X_{\scriptscriptstyle\mathrm{\leftarrow}}(t)
\end{equation}
in terms of the two correlators:
\begin{eqnarray}\label{eq:X} X_{\scriptscriptstyle\mathrm{\rightarrow}}(t) \!&=& \!\langle A^{\dagger}(t)A(0) \rangle/\hbar^2\nonumber\\\ X_{\scriptscriptstyle\mathrm{\leftarrow}}(t) \!&= &\!\langle A(0)A^{\dagger}(t) \rangle/\hbar^2
\end{eqnarray}
that form the two building blocks for time dependent transport \cite{ines_cond_mat,ines_PRB_2019}. 
One can check that the ATE link applies whenever ${H}_0 $ corresponds to a low energy effective theory for incompressible hierarchy edges, and $A$ is the dominant tunneling operator of a quasiparticle species with upper and down fields  $\psi_{u(d)}(x)$ associated with a spatially localized QPC at $x=0$: \begin{equation}\label{eq:tunA}
    A = \xi\, \psi_{u}^{\dagger}(0)\psi_{d}(0),
\end{equation}
where the tunneling amplitude~$\xi$ forms the perturbative parameter. In that case, one has either
$\omega_{\scriptscriptstyle\mathrm{dc}} = e^{*}(V_{u} - V_{d})/\hbar$ accounting for the DC bias between the upper and lower edge states in a two-terminal geometry in  \ref{fig.QPC}, or, in an "anyon collider" illustrated in Fig.~\ref{fig.QPC}, $\omega_{dc}$ is more intricately related to the DC drives applied to the source QPCs (see Eq.~\eqref{eq:omegaJ_anyon}) \cite{ines_PRB_R_noise_2020}. The ATE link can then be derived from the local exchange between anyon and quasihole fields at $x=0$. For Laughlin states (see Appendix~\ref{app:braiding_TLL}) or for multiple co-propagating channels characterized by a $\bf K$ matrix with short-range interactions encoded in $H_0$, one has $\bar{\theta}=\theta$, the statistical phase associated with the dominant tunneling anyon species. This identity does not hold for counter-propagating modes; the ATE link is valid, but with a nonuniversal phase $\bar{\theta}$ sensitive to interactions.  In all these cases based on a low-energy effective theory, one has a local TLL behavior of the DC current with a scaling dimension  $\delta=\bar{\theta} \; /\pi$.

% Example continuation (you can uncomment and adapt one of the lines below):
% extend the UNEP relation to coupled Josephson junctions where dual charge--phase variables play an analogous role.
% apply this approach to non-chiral superconducting circuits exhibiting Coulomb drag effects.
% demonstrate how braiding-like statistical phases emerge in systems beyond the fractional quantum Hall regime.

%%%%%%%%% fig 1 %%%%%%%%%%%
%%%%%%%%%%%%%%%%%%%%%%%%%%%%%%%%%

Here we consider the general UNEP Hamiltonian in Eq.~\eqref{eq:hamiltonian} without specifying $H_0$ nor $A$. The nonequilibrium DC tunneling current average and noise can be expressed as:\begin{eqnarray}I({\omega_{\scriptscriptstyle\mathrm{dc}}}) &= &e^*\left[X_{\scriptscriptstyle\mathrm{\rightarrow}}(\omega_{\scriptscriptstyle\mathrm{dc}}) - X_{\scriptscriptstyle\mathrm{\leftarrow}}(\omega_{\scriptscriptstyle\mathrm{dc}})\right]\label{eq:I_X},\\
S(\omega_{\scriptscriptstyle\mathrm{dc}})& = &e^{\scriptscriptstyle\mathrm{*2}} \left[X_{\scriptscriptstyle\mathrm{\rightarrow}}(\omega_{\scriptscriptstyle\mathrm{dc}}) + X_{\scriptscriptstyle\mathrm{\leftarrow}}(\omega_{\scriptscriptstyle\mathrm{dc}})\right]\label{eq:S_X}.
    \end{eqnarray}
The two correlators $X_{\scriptscriptstyle\mathrm{\rightarrow}}(\omega_{\scriptscriptstyle\mathrm{dc}}),X_{\scriptscriptstyle\mathrm{\leftarrow}}(\omega_{\scriptscriptstyle\mathrm{dc}})$, and consequently the two observables $I({\omega_{\scriptscriptstyle\mathrm{dc}}})$ and $S({\omega_{\scriptscriptstyle\mathrm{dc}}})$,
generally independent in absence of inversion symmetry or detailed balance, are now constrained by the ATE in Eq.~\eqref{eq:braiding_X}. Therefore, one can single out the retarded correlator as the unique building block
\begin{eqnarray}\label{eq:XR}
X^{R}(t) = \Theta(t)\left[X_{\scriptscriptstyle\mathrm{\rightarrow}}(t) - X_{\scriptscriptstyle\mathrm{\leftarrow}}(t)\right],
\end{eqnarray}
where $\Theta(t)$ is the Heaviside function. 
Taking the Fourier transform of the above expression, one obtains
\begin{eqnarray}
X^{R}(\omega)= i \int \frac{d\omega^{\prime}}{2\pi}
\frac{X_{\scriptscriptstyle\mathrm{\rightarrow}}(\omega^{\prime})
- X_{\scriptscriptstyle\mathrm{\leftarrow}}(\omega^{\prime})}
{\omega-\omega^{\prime}+i0}.
\end{eqnarray}
It is already established that the DC current is given by \cite{ines_eugene,ines_PRB_2019}
\begin{equation}\label{eq:dc_currentXR}
I(\omega_{\scriptscriptstyle\mathrm{dc}}) =
2 e^* \Re\!\left[X^R(\omega=\omega_{\scriptscriptstyle\mathrm{dc}})\right].
\end{equation}
Due to Eq.~\eqref{eq:braiding_X}, it has been demonstrated that $X^R$ also determines the DC tunneling shot-noise. The underlying trick consists of writing, using Eq.~\eqref{eq:S_X}, 
\begin{align}
S(\omega_{\scriptscriptstyle\mathrm{dc}})/e^{*2} 
&= \sum_{\epsilon=\pm} \left(1 + e^{2i\epsilon \bar{\theta}}\right) X_{\scriptscriptstyle\mathrm{\rightarrow}}^{\epsilon}(\omega_{\scriptscriptstyle\mathrm{dc}}) \\
&= (1 + e^{2i\bar{\theta}}) \left[ X_{\scriptscriptstyle\mathrm{\rightarrow}}^{+}(\omega_{\scriptscriptstyle\mathrm{dc}}) + e^{-2i\bar{\theta}} X_{\scriptscriptstyle\mathrm{\rightarrow}}^{-}(\omega_{\scriptscriptstyle\mathrm{dc}}) \right],
\end{align}
where $$X_{\scriptscriptstyle\rm{\rightarrow}}^{\epsilon}(\omega)=\int dt \; \Theta(\epsilon t) \; e^{i\omega t} X_{\scriptscriptstyle\rm{\rightarrow}}(t).$$

On the other hand, one writes, using Eq.\eqref{eq:XR}
and the property $X_{\scriptscriptstyle\mathrm{\rightarrow}}^{*}(t) = X_{\scriptscriptstyle\mathrm{\rightarrow}}(-t)$,
\begin{multline}
2i\,\Im X^{R}(\omega_{\scriptscriptstyle\mathrm{dc}})= \sum_{\epsilon=\pm} \epsilon \left(1 - e^{2i\epsilon\bar{\theta}}\right) X_{\scriptscriptstyle\mathrm{\rightarrow}}^{\epsilon}(\omega_{\scriptscriptstyle\mathrm{dc}}) \\
= (1 - e^{2i\bar{\theta}}) \left[ X_{\scriptscriptstyle\mathrm{\rightarrow}}^{+}(\omega_{\scriptscriptstyle\mathrm{dc}}) + e^{-2i\bar{\theta}} X_{\scriptscriptstyle\mathrm{\rightarrow}}^{-}(\omega_{\scriptscriptstyle\mathrm{dc}}) \right].
\end{multline}
 This provides the identity \cite{ines_statistics_2025}: 
\begin{eqnarray}\label{eq:S_FDT_braiding}
S(\omega_{\scriptscriptstyle\mathrm{dc}}) =-2e^{*2}\cot\bar{\theta}  \;\; \Im [X^R(\omega_{\scriptscriptstyle\mathrm{dc}})]
\end{eqnarray} 
for arbitrary DC frequencies $\omega_{\scriptscriptstyle\mathrm{dc}}$ within the UNEP validity range. Since $X^R$ is a response function even for nonequilibrium initial states \cite{ines_proc_2009,ines_philippe}, this forms a fundamental ATE nonequilibrium fluctuation dissipation relation. It has been shown explicitly to hold for TLL type behavior, either in a two-terminal or a collider type-geometry at zero temperature. It opens several routes to extract the ATE phase $\bar{\theta}$ from time-dependent transport, up to the usual indeterminacy mod $\pi$. In the case of fermions, $\bar{\theta}=\pi$, the divergence is compensated by the dependence of $X^R(\omega_{\scriptscriptstyle\mathrm{dc}})$ on $\bar{\theta}$.  

Applying the Kramers-Kr\"onig relations to the real and imaginary parts of $X^R$,
we subtract the zero-frequency noise $S(0)$ to ensure the convergence of the integral.
The resulting compact forms for both equations read:
 \begin{multline}
S(\omega_{dc})-S(0)
\\= -e^{*}\omega_{dc}\cot\bar{\theta}\;
{\rm P.V.}\!\!\int \frac{d\omega_{dc}'}{\pi}\,
\frac{I(\omega_{dc}')}{\omega_{dc}'(\omega_{dc}-\omega_{dc}')},
\label{eq:equation_current_noise_simple_excess}
\end{multline}
\begin{eqnarray}
e^{*}(I(\omega_{\scriptscriptstyle\mathrm{dc}})-I(0))
= \,\omega_{\scriptscriptstyle\mathrm{dc}}\tan\bar{\theta}\;
{\mathrm P.V.}\!\!\int \frac{d\omega'_{\scriptscriptstyle\mathrm{dc}}}{\pi}\,
\frac{S(\omega_{\scriptscriptstyle\mathrm{dc}}')}{\omega_{\scriptscriptstyle\mathrm{dc}}'(\omega_{\scriptscriptstyle\mathrm{dc}}-\omega_{\scriptscriptstyle\mathrm{dc}}')}.
\label{eq:equation_noise_current_excess}
\end{eqnarray}
Here $\mathrm{P.V.}$ refers to the Principal Value. Notice that for a general initial nonequilibrium distribution, we do not necessarily have $I(0)=0$ and $S(0)$ still represents a nonequilibrium noise  (see~\cite{ines_PRB_2019,ines_PRB_R_noise_2020}).

Whenever the integrals converge, these relations provide practical means to extract~$\tan\bar{\theta}$ from
measurements of either the DC current or the DC noise over a sufficiently broad frequency range~$\omega_{\scriptscriptstyle\mathrm{dc}}'$, typically centered around the external DC frequency~$\omega_{\scriptscriptstyle\mathrm{dc}}$. Although the integrand may depend on~$\bar{\theta}$, the prefactor~$\tan\bar{\theta}$ is thereby isolated. These relations apply to arbitrary stationary nonequilibrium distributions, including the "anyon collider" geometry and setups with a temperature bias. Notice that the first integral equation has been written in Ref.~\cite{ines_statistics_2025} in terms of the symmetric parts of DC current and noise in order to improve the convergence of the integral, while we keep their full form while solving this equation. 
It is useful to distinguish the role of inversion symmetry (thus an odd DC current, and particle-hole symmetry for independent electrons),  from the ATE constraint. The latter only fixes the exchange monodromy of the correlators, while inversion symmetry relates the forward and backward correlators through $X_{\scriptscriptstyle\mathrm{\rightarrow}}(t)=X_{\scriptscriptstyle\mathrm{\leftarrow}}(-t)=X(t)$. As shown in Appendix~\ref{ATE+inversion}, ATE, together with inversion symmetry, still allows for a correlator $X(t)$ defined up to an arbitrary even function of $t$, and therefore does not uniquely determine the DC transport. The uniqueness of the TLL solution emerges only after detailed balance is imposed, as will be shown in the next section.

\subsection{Comparison between ATE and braiding in the time-domain}

Let us now clarify the relation between the ATE constraint in Eq.~\eqref{eq:braiding_X} and time-domain braiding \cite{fractional_statistics_theory_Sim_Nat_comm_2016,Sim_non_abelian_NATcomm_2022,fractional_satistics_zhang_gefen_PRL_2025} using the interference picture of Ref.~\cite{feve_anyons_martin_Science_2025} for Laughlin states. For hierarchical states, we assume the injected anyon species is equivalent to the one that  dominates tunneling at the QPC. A tunneling rate at the analyzer QPC results from the interference between events occurring at two different times, $t$ and $t'$. If an
injected anyon pulse crosses the QPC during this interval, the two histories differ by the temporal ordering of the injected excitation and the quasiparticle-quasihole excitation created at the QPC. More explicitly, the injected pulse at a position $x_{\mathrm{inj}}$ is created by the anyon
operator $\Psi^{\dagger}(x_{\mathrm{inj}})$, whereas
tunneling at the central QPC is described by \eqref{eq:tunA}.
The tunneling rates involve correlators of $A(t)$ evaluated
in the state created by $\Psi^{\dagger}$.
Commuting the injected-anyon operator with the anyon fields
at the QPC produces the relative statistical factor
\begin{equation}
e^{\pm 2i\theta N(t',t)},
\qquad
N(t',t)=\frac{1}{e^*}\int_{t'}^{t}d\tau\,
I_{\rm inj}(\tau),
\end{equation}
where $I_{\mathrm{inj}}(\tau)$ is the time-dependent electrical current carried by the injected anyon pulse through the QPC and the sign depends on the tunneling direction. This accumulated phase is generally time dependent and therefore also reflects the injection protocol. Notice that the full braiding phase in
Ref.~\cite{feve_anyons_martin_Science_2025} corresponds to $2\theta$ in our convention.

Although the accumulated phase generally differs from the local ATE phase, both originate from the same anyonic exchange algebra. For hierarchical states described by a low-energy effective theory, ATE is a necessary but not sufficient condition for time-domain braiding, which, when realized, is governed by the same phase $\bar{\theta}$. Indeed, propagation over a finite distance generally spreads an injected anyon into $N$ nonuniversal components associated with the $N$ proper modes, each acquiring its own phase through time-domain braiding with anyon--anti-anyon pairs created at the QPC \cite{ines_statistics_2025,fractional_statistics_christian_splatt_FQHE_2026}. These phases collectively recover $\bar{\theta}$, but their temporal separation prevents a single braiding event from probing it directly. The standard time-domain braiding picture is recovered only when the modes have equal velocities or when injection occurs sufficiently close to the QPC to suppress temporal spreading \cite{oreg_PRL_FQHE_statistics_2023}, provided the QPC itself is spatially local. Under these conditions, the change in temporal ordering reduces to the local ATE relation with phase $\bar{\theta}$. But the identification of $\bar{\theta}=\theta$ can be shown only for short-range interacting co-propagating modes on each edge separately. This distinction also emphasizes that ATE in Eq.~\eqref{eq:braiding_X}, being a strictly local constraint at the QPC, is more robust against propagation effects than time-domain braiding protocols involving injection at a finite distance.

This distinction is also reflected at the level of the correlation functions. Time-domain braiding protocols identify the statistical phase by comparing a nonequilibrium correlator with its equilibrium counterpart~\cite{fractional_statistics_theory_Sim_Nat_comm_2016, Sim_non_abelian_NATcomm_2022}. By contrast, ATE directly relates the two temporal orderings of local tunneling operators within the same stationary state and therefore does not require an equilibrium reference state. It can consequently be formulated more generally, including dissipative quantum circuits exhibiting TLL-like transport, where $\bar{\theta}$ may be determined by the environment rather than by genuinely anyonic excitations. This motivates us to impose ATE without specifying the Hamiltonian terms $H_0$ and $A$ in Eq.~\eqref{eq:hamiltonian} in order to investigate whether it admits solutions beyond the TLL form.

\section{Thermal distribution: Poissonian noise} 
In this section, we look for DC transport consistent with the ATE constraint when the nonequilibrium DC current and noise obey the Poisson relation, so that the integral equation in Eq.~\eqref{eq:equation_current_noise_simple_excess} reduces to a closed form for the DC current that we aim to determine. This is particularly the case when the detailed balance relation holds for the correlators in Eq.~\ref{eq:X}, which, for an initial thermal state at temperature $T$ (see Appendix~\ref{app:detailed_balance}), reads:
\begin{equation}\label{eq:detailed}X_{\scriptscriptstyle\mathrm{\rightarrow}}(\omega)=e^{ \omega/\omega_{\scriptscriptstyle\rm th}}X_{\scriptscriptstyle\mathrm{\leftarrow}}(\omega),\;\;\; \omega_{\scriptscriptstyle\rm th }=k_B T/\hbar.\end{equation} Then, using Eqs.~\eqref{eq:I_X},~\eqref{eq:S_X}, the UNEP theory generalizes the DC non-equilibrium fluctuation dissipation relation \cite{ines_PRB_R_noise_2020,ines_cond_mat} beyond  Ref.\cite{levitov_reznikov}:
\begin{eqnarray}\label{FDT}
S(\omega_{\scriptscriptstyle\mathrm{dc}})
= e^{*}\coth\!\left(\frac{\omega_{\scriptscriptstyle\mathrm{dc}}}{2\omega_{\scriptscriptstyle\rm th}}\right)
I(\omega_{\scriptscriptstyle\mathrm{dc}}).
\end{eqnarray}
Therefore, one has the equilibrium FDT for the zero-frequency noise:\begin{equation}
    S(0)=2e^*\omega_{\scriptscriptstyle\rm th}G(0)\label{eq:equ_S}
\end{equation} where the rescaled linear conductance, having the dimension of an electric charge, is defined as
 \begin{eqnarray}
G(\omega_{\scriptscriptstyle\rm dc })=d I(\omega_{\scriptscriptstyle\mathrm{dc}})/d \omega_{\scriptscriptstyle\mathrm{dc}}.
\end{eqnarray}
The present analysis holds nonetheless as soon as Eq.~\eqref{FDT} holds, regardless of the physical meaning of $\omega_{\rm th}$.
Letting $J(0)=G(0)/{e^*} $, and starting from Eq.~(\ref{eq:equation_current_noise_simple_excess}), we introduce the dimensionless functions:
\begin{equation}J(\omega)=I(\omega_{\scriptscriptstyle\mathrm{dc}}=\omega)/(e^*\omega) ;\;\; F(\zeta)=J(2\omega_{\scriptscriptstyle\rm th} \zeta)/J(0), \end{equation}
of the scaled variable $\zeta=\omega/2\omega_{\scriptscriptstyle\rm th}$. Substituting this together with Eq.~\eqref{eq:equ_S} and Eq.~\eqref{FDT} where we replace $\omega=\omega_{\scriptscriptstyle\rm dc}$ 
%\begin{eqnarray}
%S(\omega)=e^*\coth\zeta\,I(\omega),
%S(0)=2e^*\omega_{\rm th}G(0),
%\end{eqnarray}
into Eq.~\eqref{eq:equation_current_noise_simple_excess}, and changing variables $\omega'=2\omega_{\rm th}\zeta'$, 
we obtain:
\begin{eqnarray}
(\coth\zeta)\,F(\zeta)-\frac1\zeta
=
-\cot\bar\theta\,
{\rm P.V.}\!\int
\frac{d\zeta'}{\pi}
\frac{F(\zeta')}{\zeta-\zeta'}.
\end{eqnarray}

At small frequencies, the principal-value integral vanishes since $J(\omega)$ is even, leaving only the leading contribution proportional to $J(0)$. By using the Sokhotski-Plemelj identity, we arrive at the final form of the equation
\begin{multline}\label{eq:equation_current_noise_residue}
(\coth \zeta + i\cot\bar{\theta})\, F(\zeta)
\\
= \frac{1}{\zeta}
-\cot \bar{\theta} \int \frac{F(\zeta^{\prime})}{\zeta - \zeta^{\prime} + i0^{+}} \frac{d\zeta^{\prime}}{\pi}.
\end{multline}
Both ATE and detailed balance constraints are therefore encoded in this functional integral equation for $F$. In fact, since $\cot\bar{\theta}$ defines $\bar{\theta}$ mod $\pi$, we adopt from now on the corresponding phase restricted to $[0,\pi]$, keeping the same notation $\bar{\theta}$ for simplicity. 

\subsection{Wiener-Hopf (W-H) method}
To find the general solution of the Eq.~(\ref{eq:equation_current_noise_residue}) we employ the W-H technique \cite{noble1958wiener}. We first briefly outline the formal W-H procedure in a general setting, introducing the multiplicative factorization of the kernel together with the additive decomposition of the inhomogeneous term. We then construct these decompositions explicitly for the kernels arising in the present problem.
The first formal step consists of decomposing the function $F(\zeta)$ as:
\begin{equation}\label{eq:F}
    F(\zeta)=F_{+}(\zeta)+F_{-}(\zeta)
\end{equation} where $F_{+}({\zeta})$ (resp. $F_{-}({\zeta})$) is an analytic function in the upper (resp. lower) half-plane. 
In this representation, the equation takes the standard W-H form,
\begin{eqnarray}\label{W-H}
F_{-}(\zeta)+K(\zeta)F_{+}(\zeta)=L(\zeta),
\end{eqnarray}
with
\begin{eqnarray}\label{eq:KL}
K(\zeta)=\frac{\coth \zeta-i\cot\bar{\theta}}{\coth \zeta+i\cot\bar{\theta}}, \;
L(\zeta)=\frac{1}{\zeta[\coth \zeta+i\cot \bar{\theta}]}.
\end{eqnarray}

The second formal step consists of a multiplicative factorization of the kernel \begin{equation}
    K(\zeta)=K_{+}(\zeta)K_{-}(\zeta)\label{eq:K}
\end{equation} into two functions analytic on the upper or lower plane. This factorization follows from the analytical properties required in the W–H method. The zeros and poles of $K(\zeta)$ are located at $(\zeta=\mp i\bar\theta+i\pi n)$, $(n\in\mathbb{Z})$, implying that $K(\zeta)$ is analytic and non-vanishing in the strip $|\operatorname{Im}\zeta|<\min(\bar\theta,\pi-\bar\theta)$.The explicit construction below establishes the existence of the factorization, which is unique up to multiplicative normalization constants within the class of analytic, nonvanishing factors with the prescribed algebraic asymptotics \cite{noble1958wiener}.
We then divide Eq.~(\ref{W-H}) by $K_{-}(\zeta)$ and write the additive decomposition of the corresponding inhomogeneous term, \begin{equation}\label{eq:R}
    R(\zeta)=L(\zeta)/K_{-}(\zeta)=R_{+}(\zeta)+R_{-}(\zeta),
\end{equation} which will be carried out explicitly below. Using Liouville’s theorem \cite{noble1958wiener}, one obtains
\begin{eqnarray}\label{W-H_form}
F_{+}(\zeta)K_{+}(\zeta)-R_{+}(\zeta)=R_{-}(\zeta)-\frac{F_{-}(\zeta)}{K_{-}(\zeta)}=E(\zeta),
\end{eqnarray}
where $E(\zeta)$ is an entire analytic function. 
If $E(\zeta)$ remains bounded at infinity, Liouville’s theorem implies that it must be constant. 
Moreover, if $E(\zeta\to\infty)\to 0$, then $E(\zeta)=0$ for all $\zeta$. 
 This is indeed the case, as shown in Appendix~\ref{app:asymptotics}.
Consequently, the general W-H solution reads:
\begin{eqnarray}
F(\zeta)=\frac{R_{+}(\zeta)}{K_{+}(\zeta)}+R_{-}(\zeta)K_{-}(\zeta).
\end{eqnarray}

Let us now achieve the multiplicative decomposition in Eq.~\eqref{eq:K}. For that, we express the kernel $K(\zeta)$ in Eq.~\eqref{eq:KL} in terms of Gamma functions, thus allowing an explicit factorization:
\begin{eqnarray}
K_{+}(\zeta)=\frac{\Gamma\left[1-\frac{i \zeta}{\pi}-\frac{\bar{\theta}}{\pi}\right]}{\Gamma\left[-\frac{i \zeta}{\pi}+\frac{\bar{\theta}}{\pi}\right]}, \;\; 
K_{-}(\zeta)=\frac{\Gamma\left[\frac{i\zeta}{\pi}+\frac{\bar{\theta}}{\pi}\right]}{\Gamma\left[1+\frac{i\zeta}{\pi}-\frac{\bar{\theta}}{\pi}\right]},
\end{eqnarray}
with an asymptotic behavior $|K_{\pm}(\zeta)|\sim \zeta^{\pm(1-2\bar{\theta}/\pi)}$.
\\
This allows us to express $R(\zeta)$, the inhomogeneous part of the equation, given by Eq.~\eqref{eq:R} once we inject the above function $K_{-}(\zeta)$ together with $L(\zeta)$  in Eq.~\eqref{eq:KL}: 
\begin{eqnarray}\label{R_function}
R(\zeta)=\frac{\sinh \zeta \sin\bar{\theta}}{\pi \zeta}\left|\Gamma\!\left[1-\frac{\bar{\theta}}{\pi}+\frac{i\zeta}{\pi}\right]\right|^{2}.
\end{eqnarray}
We now determine the functions $R_\pm$ that give the additive decomposition of $R(\zeta)$ in Eq.\eqref{eq:R}. For that, we note that if $R(\zeta)$ has simple poles $p_{n}$ ($q_{m}$) in the lower (upper) half-plane with residues $a_{n}$ ($b_{m}$), then
\begin{eqnarray}\label{contour_integrals}
R_{+}(\zeta)=\frac{1}{2\pi i}\!\int_{\Gamma_{-}}\!\frac{R(t)\,dt}{t-\zeta}=\sum^{+\infty}_{n=0}\frac{a_{n}}{p_{n}-\zeta},\\
R_{-}(\zeta)=-\frac{1}{2\pi i}\!\int_{\Gamma_{+}}\!\frac{R(t)\,dt}{t-\zeta}=-\sum^{+\infty}_{m=0}\frac{b_{m}}{q_{m}-\zeta}.\label{contour_integrals_2}
\end{eqnarray}
Here, $\zeta$ lies in the analytic strip of $R(\zeta)$,
$|\operatorname{Im}\zeta|<\pi-\bar\theta$, and along the real axis $R(\zeta)\sim \sin \bar{\theta}\,\pi^{2\bar{\theta}/\pi-1}|\zeta|^{-2\bar{\theta}/\pi}$ as $|\zeta|\to\infty$. The contours $\Gamma_{\pm}$ enclose the poles in the upper and lower half-planes, respectively. Performing the contour integration in Eqs.~\eqref{contour_integrals},~\eqref{contour_integrals_2}, we find
\begin{eqnarray}
R_{+}(\zeta)=-\frac{i}{2\zeta}\tan\bar{\theta}
\!\left[\!
\frac{\Gamma\left[1-\frac{\bar{\theta}}{\pi}\right]}{\Gamma\left[\frac{\bar{\theta}}{\pi}\right]}-
\frac{\Gamma\left[1-\frac{\bar{\theta}}{\pi}-\frac{i\zeta}{\pi}\right]}{\Gamma\left[\frac{\bar{\theta}}{\pi}-\frac{i\zeta}{\pi}\right]}
\!\right],\\
R_{-}(\zeta)=\frac{i}{2\zeta}\tan\bar{\theta}
\!\left[\!
\frac{\Gamma\left[1-\frac{\bar{\theta}}{\pi}\right]}{\Gamma\left[\frac{\bar{\theta}}{\pi}\right]}-
\frac{\Gamma\left[1-\frac{\bar{\theta}}{\pi}+\frac{i \zeta}{\pi}\right]}{\Gamma\left[\frac{\bar{\theta}}{\pi}+\frac{i\zeta}{\pi}\right]}
\!\right].
\end{eqnarray}
It is straightforward to verify that the leading terms in $R_{\pm}(\zeta)$ cancel the pole at $\zeta=0$ on the real axis.
\\
Finally, from Eq.~(\ref{W-H_form}) we obtain
\begin{eqnarray}\label{LL_solution}
F_{+}(\zeta)=-\frac{i}{2\zeta}\tan\bar{\theta}
\left[
\frac{\Gamma\left[1-\frac{\bar{\theta}}{\pi}\right]}{\Gamma\left[\frac{\bar{\theta}}{\pi}\right]}
\frac{\Gamma\left[\frac{\bar{\theta}}{\pi}-\frac{i\zeta}{\pi}\right]}{\Gamma\left[1-\frac{i\zeta}{\pi}-\frac{\bar{\theta}}{\pi}\right]}
-1
\right],
\end{eqnarray}
with the property $F_{-}(\zeta)=F^{*}_{+}(\zeta)$. 
The resulting expression for the full solution in Eq.~\eqref{eq:F} reads:
\begin{eqnarray}\label{TLL}
F(\zeta)=F_{+}(\zeta)+F_{-}(\zeta)
=\frac{\sinh \zeta}{\zeta}\frac{\left|\Gamma\left[\frac{\bar{\theta}}{\pi}+\frac{i \zeta}{\pi}\right]\right|^{2}}{\Gamma\left[\frac{\bar{\theta}}{\pi}\right]^{2}},
\end{eqnarray}
which indeed satisfies $F(0)=1$. By rewriting this solution in terms of the tunneling current, we arrive at:
\begin{multline}\label{answer}
 I(\omega_{\scriptscriptstyle\mathrm{dc}})= \omega_{\scriptscriptstyle\mathrm{dc}}G(0)F(\omega_{\scriptscriptstyle\mathrm{dc}}/2\omega_{\scriptscriptstyle\rm th})\\= 2\omega_{\scriptscriptstyle\rm th}G(0) \sinh\left(\frac{\omega_{\scriptscriptstyle\mathrm{dc}}}{2\omega_{\scriptscriptstyle\rm th}}  \right) \frac{\left|\Gamma\left[\frac{\bar{\theta}}{\pi}+\frac{i\omega_{\scriptscriptstyle\mathrm{dc}}}{2\pi \omega_{\scriptscriptstyle\rm th}}\right]\right|^{2}}{\Gamma\left[\frac{\bar{\theta}}{\pi}\right]^{2}}.  
\end{multline}
This solution reproduces the standard TLL result for the tunneling current through a spatially local QPC, with a scaling dimension $0<\delta<1$ corresponding to the angle $\bar{\theta}$ now restricted to $[0,\pi]$. The restriction $0<\delta<1$ arises from the mathematical implementation of the W-H analysis rather than from a fundamental limitation of the ATE constraint itself, which is equivalent for all angles $\bar{\theta} \!\!\mod \pi$; indeed, the resulting analytical expression for the tunneling current~\eqref{answer} can be extended to $\bar{\theta}/\pi=\delta>1$ by analytic continuation.

We have also verified that the second Kramers-Kr\"onig relation in Eq.~(\ref{eq:equation_noise_current_excess}) is satisfied. However, there remains an indeterminacy in $G(0)$ due to the homogeneity of the ATE integral equation in Eq.~\eqref{eq:equation_current_noise_simple_excess} with respect to this quantity. As explained in Appendix~\ref{app:spectral}, the nonequilibrium current in the regime $\omega_{\scriptscriptstyle\mathrm dc} \gg \omega_{\scriptscriptstyle\rm th}$ depends solely on $\omega_{\scriptscriptstyle\mathrm dc}$, which fixes the power-law behavior $G(0) \propto \omega_{\scriptscriptstyle\rm th}^{2\delta - 2}$, with a prefactor determined by the microscopic details of the model which, in the case $A$ has the form in Eq.\eqref{eq:tunA}, is proportional to the reflection coefficient at the QPC  $|\xi|^2$ in Eq.~\eqref{eq:tunA}. 

One may nevertheless wonder whether the uniqueness of the solution is an artifact of rewriting the problem in terms of a Kramers-Kr\"onig relation. Indeed, the latter requires the convergence of the integral in Eq.~\eqref{eq:equation_noise_current_excess} and is therefore more restrictive than the bare ATE constraint in Eq.~\eqref{eq:braiding_X}. It is thus important to solve the ATE constraint directly, as done in Appendix~\ref{ATE+Detailed balance}. We show that once combined with detailed balance, the ATE constraint again yields a unique solution corresponding to the TLL expression for the DC current. 

This emergence occurs only upon imposing both the ATE constraint and the detailed-balance condition specific to initial thermal states. Their combination uniquely yields the TLL behavior, naturally characterized by a scaling dimension $\delta$. The identification $\pi\delta=\bar{\theta}\;(\mathrm{mod}\;\pi)$ is therefore not a result deduced from the analysis of a low-energy effective model for a local QPC \cite{ines_statistics_2025}. In that framework, for short-range intermode interactions, one obtains the stronger identity $\pi\delta=\bar{\theta}$, without the modulo-$\pi$ ambiguity, and furthermore $\pi\delta=\bar{\theta}=\theta$ when all modes have the same chirality. Here, instead, the relation $\pi\delta=\bar{\theta}\;(\mathrm{mod}\;\pi)$ emerges directly and solely from combining the ATE constraint with detailed balance.

 We stress that the TLL solution for the DC tunneling current does not require the edge Hamiltonian $H_0$  to be formed by a free chiral TLL model, as this is not necessary to derive the ATE link. For instance, the latter holds for multiple interacting channels within the effective edge theory, where the robustness of the TLL behavior reflects the locality of the ATE constraint at the QPC  \cite{ines_statistics_2025}.  It can also be checked in a nonchiral system with a local impurity, in a coherent conductor connected to an ohmic environment \cite{ines_saleur}. We can also check that the ATE link holds in a toy model consisting of an edge coupled to a localized two-level system (see Appendix~\ref{TLS}). 

\section{Superpoissonian noise: Application to the "Anyon Collider"}

A particularly relevant arena for implementing ATE protocols is provided by "anyon collider" geometries Fig.~\ref{fig.collider}, where quasiparticles emitted from independent QPC sources are partitioned at a central QPC. Beyond thermal equilibrium, the integral relations \eqref{eq:equation_current_noise_simple_excess} and \eqref{eq:equation_noise_current_excess} remain valid for arbitrary stationary nonequilibrium distributions within UNEP theory. In such cases, the structure of the response function
is not restricted to a TLL form, and the ATE constraint does not necessarily imply the existence of a scaling behavior of the DC current with respect to voltage and temperature, as additional energy scales are expected to enter.

\subsection{Prescription for the QPC sources extended to hierarchical edge states}
\begin{figure}[t!]
\centering
\includegraphics[width= 0.99 \linewidth]{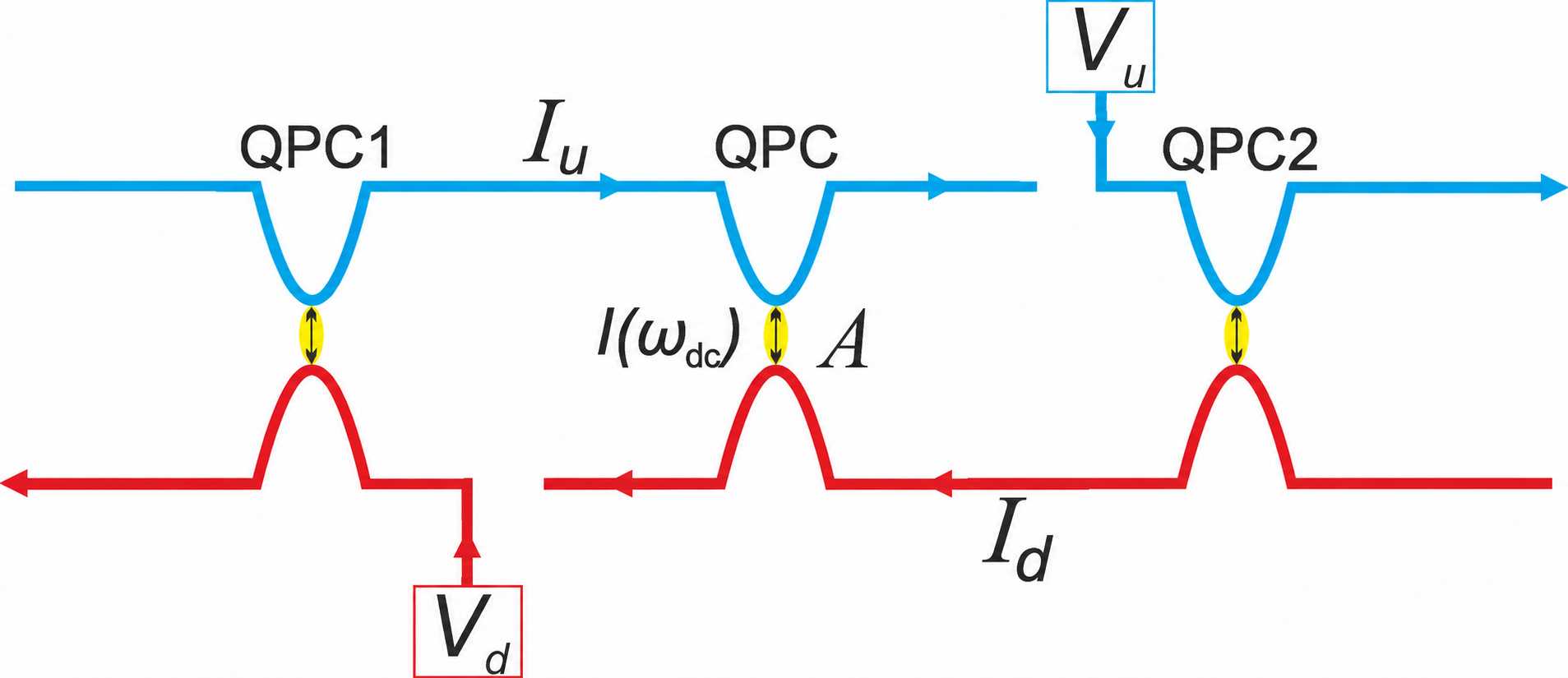}
\caption{An anyon collider setup in the FQHE. Two QPCs, subject to
DC-biases $V_{u(d)}$, inject anyons into the upper/down edges that tunnel at the central QPC.
\label{fig.collider}}
\end{figure}

We denote by $I_u$ and $I_d$ the weak injected currents along the upper and lower edges (see Fig.~\ref{fig.collider}), respectively, both taken as positive towards the analyzer QPC, and we define:
\begin{equation}
I_+=I_u+I_d,
\qquad
I_-=I_u-I_d.
\end{equation}
In order to incorporate these currents in the nonequilibrium correlators as presented in Ref.~\cite{fractional_statistics_theory_2016}, we now adopt the hypothesis of a bosonized Hamiltonian for the edge states, in addition to the ATE constraint. This allows us to use the nonequilibrium bosonization formalism, initially implemented in Ref.~\cite{ines_epj} through the equations of motion for bosonic fields whose boundary conditions are provided by time-dependent sources; this method accounts for an arbitrary nonequilibrium distribution and led to the matrix scattering approach for magnetoplasmons.

The latter, considered in a more extended analysis in Refs.~\cite{out_of_equilibrium_bosonisation_eugene_PRL_2009,out_of_equilibrium_bosonisation_gutman_epl_2010} however restricted to the DC regime, arises from DC currents with a poissonian distribution injected by DC non-Gaussian dilute sources, such as quantum point contacts (QPCs) dictating the stationary boundary conditions. 

In Ref.~\cite{fractional_statistics_theory_2016}, which focuses on Laughlin states, a phenomenological extension to hierarchical Hall states introduced a nonuniversal renormalization parameter $\lambda$ of the injected currents, accounting for inter-channel interactions (and edge reconstruction). Through another phenomenological analysis within low energy effective theory for higher-order states characterized by a $\mathbf{K}$ matrix~\cite{oreg_PRL_FQHE_statistics_2023}, it was shown that $\lambda = \theta / \pi$, provided  the dispersion of the anyon is reduced, or the modes are chiral with identical velocities $v$. This corresponds to a specific interaction matrix between edge channels $\mathbf{U} $ that obeys $\mathbf{K}^{-1}\mathbf{U}=v\;\mathbf{Id}$.

However, the identity $\lambda = \theta / \pi$ in hierarchical Hall states has more recently been derived through a more rigorous argument \cite{ines_statistics_2025} based on the matrix scattering approach for magnetoplasmons, expressing the solution for time-dependent bosonic fields in terms of arbitrary time-dependent boundary conditions on the injected current components. The argument uses current conservation alone and, as such, generalizes to cases involving inter-channel interactions, non-diagonal $\mathbf{K}$ matrices, and even arbitrary quadratic Hamiltonians with respect to the channel bosonic fields. This allows us to adopt in the more general framework of the hierarchical interacting edge states the source-induced parameters \cite{fractional_statistics_theory_2016,ines_PRB_R_noise_2020,fractional_statistics_rosenow_halperin_proceedings_2025}:
\begin{eqnarray}
\label{eq:omegaJ_anyon}
\omega_+ =2\sin^2\theta\frac{I_+}{e^*} \quad \omega_- = \sin(2\theta)\frac{I_-}{e^*}.
\end{eqnarray}
The balanced configuration corresponds to $I_-=0$, thus $\omega_{-}=0$. These two frequencies dictate the correspondence between the nonequilibrium correlation functions of the tunneling operator in the presence of source QPCs and their reference counterparts in the absence of these sources~\cite{fractional_statistics_theory_2016}:
\begin{eqnarray}
\langle A(t)A^{\dagger}(0)\rangle_{\scriptscriptstyle\mathrm{sources}}=e^{-\omega_{+}|t|}e^{-i\omega_{-}t}\langle  A(t)A^{\dagger}(0) \rangle    
\label{coll}
\end{eqnarray}
In full generality,  average on the right side, taken in the absence of the source QPCs, does not require an explicit
specification of an equilibrium reference density matrix. In particular, the latter may
be nonthermal and describe a stationary nonequilibrium state, for example when the edge channels are not fully equilibrated or when temperature biases are present between
different channels or between the upper and lower edges at the central QPC. 
$\omega_{-}$ and $\omega_{+}$ play fundamentally different roles. The frequency $\omega_{-}$ corresponds to the effective DC drive generated by the imbalance between the two source QPCs and therefore enters the UNEP theory exactly as the DC frequency in the two-terminal geometry \cite{ines_PRB_R_noise_2020}. In the following, we therefore identify $\omega_{dc}\equiv \omega_{-}$~\cite{fractional_statistics_theory_2016} as it plays the role of the nonequilibrium DC drive in Eq.~\eqref{eq:hamiltonian}. By contrast, $\omega_{+}$ is an additional parameter characterizing the nonequilibrium states generated by the source QPCs and controlling the energy broadening of the injected anyon distribution, thereby introducing an extra energy scale into the problem. 
\subsection{Tunneling current and noise for a reference nonequilibrium distribution }
In order to align with the previous analysis within the UNEP framework, we need to omit $e^{-i\omega_{-}t}$  in the definition of the nonequilibrium correlators in Eqs.~\eqref{eq:X}, as they are defined in the absence of $\omega_{dc}= \omega_{-}$ but encode $\omega_{+}$. Adding the subscript "coll"  to all correlators and observables, we have: 
\begin{equation}
\label{eq:X_coll}
\begin{aligned}
X_{\scriptscriptstyle\rightarrow,\mathrm{coll}}(t)
&=
e^{-i\omega_{-}t}
\langle A^{\dagger}(t)A(0)\rangle_{\mathrm{sources}}/\hbar^2
=
e^{-\omega_{+}|t|}
X_{\scriptscriptstyle\rightarrow}(t),
\\
X_{\scriptscriptstyle\leftarrow,\mathrm{coll}}(t)
&=
e^{-i\omega_{-}t}
\langle A(0)A^{\dagger}(t)\rangle_{\mathrm{sources}}/\hbar^2
=
e^{-\omega_{+}|t|}
X_{\scriptscriptstyle\leftarrow}(t).
\end{aligned}
\end{equation}
where $X_{\scriptscriptstyle\mathrm{\rightarrow}}(t),X_{\scriptscriptstyle\mathrm{\leftarrow}}(t)$ are the reference correlators in absence of the QPC sources. Let us now impose the ATE constraint given in Eq.~\eqref{eq:braiding_X} on these two correlators, thus $X_{\scriptscriptstyle\mathrm{\rightarrow},coll}(t) =e^{-2 i \bar{\theta} \, \text{sign}(t) }X_{\scriptscriptstyle\mathrm{\leftarrow},coll}(t) $. On one hand, one has exactly the same relations as in Eqs.~\eqref{eq:dc_currentXR} and \eqref{eq:S_FDT_braiding} where we introduce the two DC drive variables:
\begin{eqnarray}
I_{\scriptscriptstyle\mathrm{coll}}(\omega_{-},\omega_{+})=2e^{*}{\Re}[X_{\scriptscriptstyle\mathrm{coll}}^{R}(\omega=\omega_{-},\omega_{+})],\\
S_{\scriptscriptstyle\mathrm{coll}}(\omega_{\scriptscriptstyle\mathrm{-}},\omega_{+}) =-2e^{*2}\cot\bar{\theta}  \;\; \Im [X_{\scriptscriptstyle\mathrm{coll}}^R(\omega=\omega_{\mathrm{-}},\omega_{+})],
\end{eqnarray}
and the retarded function is now given by
\begin{eqnarray}
X_{\scriptscriptstyle\mathrm{coll}}^{R}(t)=\Theta(t)e^{-\omega_{+}t}[X_{\scriptscriptstyle\mathrm{\rightarrow}}(t)-X_{\scriptscriptstyle\mathrm{\leftarrow}}(t)].  
\end{eqnarray}
As a consequence, the fluctuation-dissipation relations~\eqref{eq:equation_current_noise_simple_excess} and \eqref{eq:equation_noise_current_excess} continue to hold with respect to the DC variable $\omega_{-}$, while $\omega_{+}$ enters only as an external nonequilibrium parameter. It is important to note that, experimentally, one generally has access to $I_-$ instead of $\omega_{\rm dc}=\omega_-$ as $\theta$ is not known in advance; thus replacing $I_{\scriptscriptstyle\mathrm{coll}}(\omega_{-},\omega_{+})=I_{\scriptscriptstyle\mathrm{coll}}(I_{-},I_{+})$ and similarly for $S_{\scriptscriptstyle\mathrm{coll}}$, we change variables in the integral equation \eqref{eq:equation_current_noise_simple_excess} that cancel the phase sensitive prefactors in $\omega_-$ in Eq.~\eqref{eq:omegaJ_anyon}:
 \begin{multline}
S_{\scriptscriptstyle\mathrm{coll}}(I_{\scriptscriptstyle\mathrm{-}},I_{+}) -S_{\scriptscriptstyle\mathrm{coll}}(0,I_{+})
\\= -e^{*}I_{-}\text{sign}(\sin(2\theta))\cot\bar{\theta}\;
{\rm P.V.}\!\!\int \frac{dI'_{-}}{\pi}\,
\frac{I_{\scriptscriptstyle\mathrm{coll}}(I_{\scriptscriptstyle\mathrm{-}}',I_{+})}{I_{-}'(I_{-}-I_{-}')}.
\label{eq:equation_current_noise_simple_excess_new}
\end{multline}
Here, one needs to vary $I_-$ while maintaining the same variable $I_+$ on both sides. Thus, for a given value of $I_+$, one has to vary $I_u$ and $I_d$ in a correlated fashion, as $2I_u=I_++I_-$ and $2I_d=I_+-I_-$.  Therefore, information on the ATE phase could be gained from the integral equations \eqref{eq:equation_current_noise_simple_excess} or \eqref{eq:equation_noise_current_excess}. But one also needs to anticipate $\text{sign}(\sin(2\theta))$; thus the situation $\theta=\bar{\theta}$ would be the easiest to handle, as then$sign(\sin(2\theta))\cot{\theta}=|\cot{\theta}|$  and this integral equation offers routes to access the universal phase $\theta$. We recall that $\theta=\bar{\theta}$  for low-energy effective theories of  interacting co-propagating edge modes.

On the other hand, the source QPCs merely modify the envelope of the correlators through the even factor $e^{-\omega_{+}|t|}$. As a consequence, the ATE constraint remains unchanged in the time domain. Therefore, the class of DC current and noise functions compatible with the ATE constraint in the "anyon collider" geometry is entirely determined by the corresponding nonequilibrium DC current $I(\omega_{-})$ and noise $S(\omega_-)$ satisfying the ATE constraint in the absence of the source QPCs. The effect of the sources is then incorporated through a convolution with an appropriate kernel: \begin{eqnarray}
 L_{\omega_{+}}(\omega)=\frac{2\omega_{+}}{\omega^{2}_{+}+\omega^{2}}. 
 \end{eqnarray}
Thus, the expressions for the DC current in \eqref{eq:I_X} and noise \eqref{eq:S_X} are generalized as follows:
\begin{multline}\label{I^{neq}}
I_{\scriptscriptstyle\mathrm{coll}}(\omega_{-},\omega_{+})=e^{*}\int \frac{d\omega}{2\pi}L_{\omega_{+}}(\omega_{-}-\omega)[X_{\scriptscriptstyle\mathrm{\rightarrow}}(\omega)-X_{\scriptscriptstyle\mathrm{\leftarrow}}(\omega)]\\=\int \frac{d\omega}{2\pi}L_{\omega_{+}}(\omega_{-}-\omega)I(\omega),
\end{multline}
\begin{multline}\label{S^{neq}}
S_{\scriptscriptstyle\mathrm{coll}}(\omega_{-},\omega_{+})=(e^{*})^{2}\int \frac{d\omega}{2\pi}L_{\omega_{+}}(\omega_{-}-\omega)[X_{\scriptscriptstyle\mathrm{\rightarrow}}(\omega)+X_{\scriptscriptstyle\mathrm{\leftarrow}}(\omega)]\\=\int \frac{d\omega}{2\pi}L_{\omega_{+}}(\omega_{-}-\omega)S(\omega),
\end{multline}
 
 \subsection{Tunneling current and noise for a thermal reference distribution} 
 
 We now consider the case where, in the absence of the source QPCs, the system is described by an equilibrium thermal density matrix. In this situation, the ATE constraint together with the detailed-balance relation of Eq.~\eqref{eq:detailed} for the two equilibrium correlators in Eq.~\eqref{eq:X} uniquely determines the latter, obeying the inversion symmetry imposed by the W-H solution:
\begin{eqnarray}
X_{\scriptscriptstyle\rightarrow}(\omega)
=
X_{\scriptscriptstyle\leftarrow}(-\omega)
=
X(\omega),
\end{eqnarray}
where we have introduced the single correlator $X$. This yields the TLL form of the DC current in Eq.~\eqref{answer} (see Appendix~\ref{ATE+Detailed balance}) with a scaling dimension $\delta=\bar{\theta}\mod\pi /\pi$ and the poissonian DC noise in Eq.~\eqref{FDT}; both observables now determine the unique solution for DC current and superpoissonian noise in Eqs.~\eqref{I^{neq}} and~\eqref{S^{neq}}.

In the "anyon collider" geometry, this inversion symmetry is preserved, but not the detailed balance, which is violated by the presence of the additional nonequilibrium scale $\omega_{+}$ that characterizes the injected nonthermal quasiparticle distribution. Nevertheless, DC transport is constrained by Eq.~\eqref{coll}, which provides a specific relation between correlation functions in the presence and absence of the source QPCs. Consequently, once the equilibrium correlator $X(\omega)$ is determined, the DC transport properties of the "anyon collider" are completely fixed.

Therefore, the uniqueness of the resulting expressions for the nonequilibrium DC current average and noise ultimately stems from the uniqueness of the thermal solution in the absence of the source QPCs. They can be directly determined through Eq.~\eqref{coll}, without the need to solve the integral equation~\eqref{eq:equation_current_noise_simple_excess_new}, which now involves both the average current and the noise.
%results obtained below should be viewed as consequences of the unique thermal TLL solution consistent with ATE constraint, combined with the non-thermal prescription of Eq.~\eqref{coll}, rather than as a new uniqueness result derived directly from the ATE constraint. 

%\ipl{It is important to emphasize the conceptual difference between the present nonequilibrium situation and the thermal case discussed in the previous sections.}

Let us now calculate the non-equilibrium current average and noise using the TLL solution in Eq.~\eqref{answer} obtained for initial thermal states. By substituting~\eqref{answer} in~\eqref{I^{neq}}, we obtain  
\begin{multline}\label{current}
I_{\scriptscriptstyle\mathrm{coll}}(\omega_-,\omega_+)
=4 |\xi|^{2} e^{*}  (2\pi \omega_{\scriptscriptstyle\rm th})^{2\bar{\theta}/\pi-1} \tau^{2\bar{\theta}/\pi}_{c}\sin \bar{\theta}
\\ \times  \Im\!\left[
B\!\left(
\bar{\theta}/\pi+\frac{\omega_+-i\omega_-}{2\pi\omega_{\scriptscriptstyle\rm th}},
\,1-2\bar{\theta}/\pi
\right)
\right].
\end{multline}
where $B(x,y)$-is the Euler beta function. Taking the zero temperature limit ($\omega_{\scriptscriptstyle\rm th}\to 0$) of this expression, we  recover the well known result~\cite{fractional_statistics_theory_2016} for the DC current with $\bar{\theta}/\pi=\delta$,
\begin{multline}\label{eq:current_coll}
I_{\scriptscriptstyle\mathrm{coll}}(\omega_{-},\omega_{+})=4 |\xi|^{2} e^{*} \tau^{2\bar{\theta}/\pi}_{c} \sin \bar{\theta} \;\Gamma[1-2\bar{\theta}/\pi] \\ \times \Im[(\omega_{+}-i\omega_{-})^{2\bar{\theta}/\pi-1}].
\end{multline}
Similarly, using the DC noise in Eq.~\eqref{FDT}, one obtains the "anyon collider" DC tunneling noise
\begin{multline}\label{eq:noise_coll}
S_{\scriptscriptstyle\mathrm{coll}}(\omega_-,\omega_+)
=
4 |\xi|^{2} e^{*2}  (2\pi \omega_{\scriptscriptstyle\rm th})^{2\bar{\theta}/\pi-1} \tau^{2\bar{\theta}/\pi}_{c} \cos \bar{\theta} 
\\ \times \Re\!\left[
B\!\left(
\bar{\theta}/\pi+\frac{\omega_+-i\omega_-}{2\pi\omega_{\scriptscriptstyle\rm th}},
\,1-2\bar{\theta}/\pi
\right)
\right],
\end{multline}
which provides the known zero temperature expression of~\cite{fractional_statistics_theory_2016}
\begin{multline}
S_{\scriptscriptstyle\mathrm{coll}}(\omega_{-},\omega_{+})=4 |\xi|^{2}e^{*2} \tau^{2\bar{\theta}/\pi}_{c} \cos \bar{\theta}\;\Gamma[1-2\bar{\theta}/\pi]  \\ \times \Re[(\omega_{+}-i\omega_{-})^{2\bar{\theta}/\pi-1}].
\end{multline}  

%%%%%%%%% fig 2 %%%%%%%%%%
%%%%%%%%%%%%%%%%%%%%%%%%%%%%%%%%%
Notice that the source-induced scale $\omega_{+}$ breaks the single-parameter scaling of the thermal two-terminal problem. The tunneling current average and noise expressions in Eqs.~\eqref{current} and~\eqref{eq:noise_coll}  at finite temperatures recover those obtained in~\cite{Sim_non_abelian_NATcomm_2022} based on a free TLL description for the edges.  By contrast, the present construction selects the TLL-type form of the thermal correlator within the physical analytic class specified above by combining ATE with thermal detailed balance. Here, hierarchical fractional edges are not assumed to be described by a free chiral TLL Hamiltonian to derive the DC current characteristics at the QPC. Nonetheless, we need a bosonized quadratic Hamiltonian, though arbitrary, that ensures the extension of the prescription in Eqs.~\eqref{coll},~\eqref{eq:omegaJ_anyon}, which is more general than a pure chiral TLL model.

\subsection{Chiral-current cross-correlations}
\label{sec:chiral-current-cross-correlations}

Experimentally, the zero-frequency cross-correlation between the 
chiral currents is more directly accessible than the tunneling noise itself. It is given by the zero-frequency Fourier transform of correlations between outgoing upper and lower edge current operators $\hat{I}_u,\hat{I}_d$, thus denoted by $S_{ud}$.
At zero temperature, the perturbative relation derived in Ref.~\cite{fractional_statistics_theory_2016} reads
\begin{align}
  S_{ud}(\omega_-,\omega_+)
  ={}&-S_{\mathrm{coll}}(\omega_-,\omega_+)
  \nonumber\\
  &+e^*\bigl(I_-\partial_{I_+}+I_+\partial_{I_-}\bigr)
  I_{\mathrm{coll}}(\omega_-,\omega_+).
\end{align}
Since the current imbalance $I_-$ introduces an effective voltage drop, the last term in this equation can be written as $\sin(2\theta)\; I_+  G_{\mathrm{coll}}(\omega_-,\omega_+)$  in terms of an effective rescaled differential conductance (having the dimension of a charge):
\begin{align}
  G_{\mathrm{coll}}(\omega_-,\omega_+)=
  \partial_{\omega_-}I_{\mathrm{coll}}(\omega_-,\omega_+),
  \label{eq:collider-differential-conductance}
\end{align}
At finite temperature, we include the mixed contribution
between the incoming thermal current fluctuations and the
analyzer tunneling current, $2e^*\omega_{\rm th}G_{\rm coll}(\omega_-,\omega_+)$,
recalling that:
\begin{equation}
  \omega_{\mathrm{th}}=\frac{k_{\mathrm B}T}{\hbar}.
\end{equation}  
For independent dilute Poissonian sources, the resulting
relation, to leading order in the QPC average tunneling current, reads

\begin{align}
S_{ud}(\omega_-,\omega_+)
={}&-S_{\mathrm{coll}}(\omega_-,\omega_+)
\nonumber\\
&+\bigl[I_+\sin(2\theta)+2e^*\omega_{\rm th}\bigr]
  G_{\mathrm{coll}}(\omega_-,\omega_+)
\nonumber\\
&+e^*I_-\partial_{I_+}
  I_{\mathrm{coll}}(\omega_-,\omega_+).
\label{eq:finite-T-output-cross-correlation}
\end{align}
The derivation relies on an independent-source prescription that does not impose a separate hierarchy between the average tunneling current at the central QPC and the injected source
currents. It requires weak tunneling at the central QPC, with higher-order corrections remaining small. The average tunneling current alone does not provide a
criterion for perturbative validity: it vanishes for
balanced injection, whereas tunneling fluctuations remain finite. The additional term
$2e^*\omega_{\rm th}\,G_{\mathrm{coll}}(\omega_-,\omega_+)$
in Eq.~\eqref{eq:finite-T-output-cross-correlation}
ensures consistency with the equilibrium
fluctuation-dissipation theorem. Indeed, in the absence
of injection, $I_+=I_-=0$, one has
$S_{\mathrm{coll}}(0,0)
=2e^*\omega_{\rm th}G_{\mathrm{coll}}(0,0)$,
so that the output cross-correlation $S_{ud}(0,0)=0$ vanishes
in the equilibrium limit.

Equation~\eqref{eq:finite-T-output-cross-correlation}
assumes that the two source QPCs act as independent dilute
Poissonian emitters, as in the DC injection protocol of the "anyon collider"
considered here. Finite temperature is accounted for in
the thermal reference correlators, while source injection
is treated as unidirectional. Outside this Poissonian regime,
the mixed source-tunneling correlations require a separate
analysis.

We now consider the balanced configuration, $I_-=0$, and adopt the 
generalized Fano factor in \cite{fractional_statistics_theory_2016}
\begin{equation}
  \mathcal F_{\omega_{\mathrm{th}}}(\omega_+)
  =\frac{S_{ud}(0,\omega_+)}
  {e^*I_+\,\partial_{I_-}I_{\mathrm{coll}}(0,\omega_+)}=\tan\theta\frac{S_{ud}(0,\omega_+)}
  {e^*\omega_+\,G_{\mathrm{coll}}(0,\omega_+)}.
  \label{eq:corrected-balanced-fano-definition}
\end{equation}
Using Eq.~\eqref{eq:omegaJ_anyon} 
the contribution of the mixed term in
Eq.~\eqref{eq:finite-T-output-cross-correlation} to the normalized quantity simplifies to
\begin{equation}
\tan\theta \frac{2e^*\omega_{\rm th} G_{\mathrm{coll}}(0,\omega_+)}
 {e^*\omega_+\,G_{\mathrm{coll}}(0,\omega_+)}
 =\frac{\tan\theta}{\pi y}.
 \label{eq:normalized-mixed-thermal-term}
\end{equation}
Here, we introduce the dimensionless ratio and a thermal current scale $I_{th}$ depending on $\theta$
\begin{equation}
y=\frac{\omega_+}{2\pi\,\omega_{\rm th}}=\frac{I_+}{I_{\mathrm{th}}},\; I_{\mathrm{th}}=e^*\frac{\pi\omega_{\rm th}}{\sin^2\theta}
\label{eq:dimensionless-temperature}
\end{equation}

Combining Eqs.~\eqref{current}, \eqref{eq:noise_coll},
\eqref{eq:finite-T-output-cross-correlation}, \eqref{eq:corrected-balanced-fano-definition} and replacing
$\bar\theta=\pi\delta$, we therefore obtain
\begin{multline}
 \mathcal F_{\omega_{\mathrm{th}}}(\omega_+)
 =1+\frac{\tan\theta}{\pi y}
\\
 +\frac{\tan\theta}{y\tan(\pi\delta)}
 \frac{1}{\,[\psi(y+\delta)-\psi(y+1-\delta)]}
  \label{eq:corrected-finite-T-fano}
\end{multline}
with $y=\omega_+/(2\pi\omega_{\mathrm{th}})$ and $\psi$ the
digamma function.  The second term in
Eq.~\eqref{eq:corrected-finite-T-fano} is precisely the contribution of
$2e^*\omega_{\rm th} G_{\mathrm{coll}}(0,\omega_+)$. The balanced generalized Fano factor in Eq.~\eqref{eq:corrected-finite-T-fano} is invariant under
$\delta\to1-\delta$ whenever $\theta$ is independent of $\delta$, in which case 
 $\mathcal F_{\omega_{\mathrm{th}}}(\omega_+)$ cannot distinguish, for example, $\delta=2/5$ from $\delta=3/5$. 

Before discussing the full temperature or injected current dependencies of the Fano factor, we first analyze its two complementary asymptotic behaviors controlled by the ratio $y$ in \eqref{eq:dimensionless-temperature}, 
 $y\gg1$ or $y\ll1$,  the low temperature or low current domains corresponding to the smallest current scale given either by $I_{th}$ or  by $I_+$ respectively.
 \subsection{Low-temperature domain}
Let us first check the zero-temperature limit, where we recover the expression obtained in
\cite{oreg_PRL_FQHE_statistics_2023,fractional_statistics_rosenow_halperin_proceedings_2025}:
\begin{equation}
 \mathcal F_0
 =\lim_{\omega_{\mathrm{th}}\ll\omega_+}
 \mathcal F_{\omega_{\mathrm{th}}}(\omega_+)
 =1-
 \frac{\tan\theta}
 {(1-2\delta)\tan(\pi\delta)},
 \label{eq:corrected-fano-zero-T}
\end{equation} 
However, there are two limitations to these previous works that are overcome by our present study of temperature effects. 
The first limitation arises from the zero-temperature assumption, which is never strictly realized experimentally: one must specify the relevant energy scale against which the unavoidable finite temperature should be compared. Here, we recover the known expression for $\mathcal F_0$ by requiring more precisely $\omega_{\rm th}\ll\omega_+$, or equivalently $I_+\gg I_{\rm th}$. However, the dilute injection required for a Poissonian distribution imposes a weak injected current, $I_+=2I_u=2I_d$, typically compared with $\nu e^2 V/h$, where $V=V_u=V_d$ is the DC voltage applied equally to each source QPC. Therefore, Eq.~\eqref{eq:corrected-fano-zero-T} cannot be reached for arbitrarily small $I_+$.

A further difficulty is that $\theta$ is not known \emph{a priori}, making the relevant scale $I_{\rm th}$ itself phase dependent. For example, for $e^*=e/5$, one finds $I_{\rm th}/(e\omega_{\rm th})\simeq0.7$ for $\theta=3\pi/5$, but $\simeq1.82$ for $\theta=\pi/5$. At $\nu=1/3$, this ratio is $\simeq1.4$, so the low-temperature condition $I_{\rm th}\ll I_+$ requires still larger injected currents.
It is therefore important to ask whether existing experiments actually reached the regime required for the zero-temperature limit in Eq.~\eqref{eq:corrected-balanced-fano-definition}. In Ref.~\cite{pierre_anyons_PRX_2023}, $T=35\,{\rm mK}$ corresponds to $e\omega_{\rm th}\simeq734\,{\rm pA}$, whereas the injected current does not exceed $80\,{\rm pA}$, where deviations from the dilute limit already occur. Thus, the low temperature regime underlying Eq.~\eqref{eq:corrected-fano-zero-T} is not reached even in the relatively favorable case $e^*=e/5$ and $\theta=3\pi/5$, for which $I_{\rm th}\simeq510\,{\rm pA}$; for $\theta=\pi/5$, the low temperature regime imposes higher injected currents, $I_+\gg I_{\rm th}\simeq1.34\,{\rm nA}$.

In Ref.~\cite{fractional_statistics_gwendal_PRX_2023}, the temperature is of the same order as in  Ref.~\cite{pierre_anyons_PRX_2023}, while $I_+$ reaches $0.2$--$0.4\,{\rm nA}$ owing to the larger voltage applied to the source QPCs. Without explicitly computing the cross-correlations and Fano factor, the lower bound on the injected current was taken to be $e^*\omega_{\rm th}$, without the phase-dependent prefactors derived here. Moreover, it is unclear whether increasing $I_+$ sufficiently to reach the low-temperature regime remains compatible with dilute injection. These considerations motivate our analysis below of the Fano factor at lower injected currents compared to $I_{th}$, a domain that is more directly accessible experimentally.
 
The second limitation of the previous works \cite{fractional_statistics_gwendal_science_2020,fractional_statistics_gwendal_PRX_2023,pierre_anyons_PRX_2023}  arises from the dependence of  $\mathcal F_0$ on two unknown parameters $\theta$ and $\delta$, leading experimental works to assume the locked case $\theta=\bar\theta=\pi\delta$ so that $\delta$ is the single unknown parameter to be determined: 
\begin{equation}
 \mathcal F_0=1-\frac{1}{1-2\delta}.
 \label{eq:locked-zero-T-fano}
\end{equation}

However, if one adopts low-energy effective theories for the edges, the universal phase $\theta$ entering {the phase-sensitive factors in} Eq.~\eqref{eq:omegaJ_anyon} can be justified for an arbitrary profile and range of the interactions between co-propagating edge modes \cite{ines_statistics_2025}, whereas the identity $\delta = \theta/\pi$ is robust only for short-range interactions or a sufficiently weak dispersion of the proper mode frequencies; in general, $\delta$ may therefore differ from $\theta/\pi$. Beyond these theories, the W-H solution allows for $\delta$ to be determined only modulo an integer relative to $\theta/\pi$, and thus does not exclude either $\delta > 1$. 

Our analysis shows that this second limitation can be overcome by examining the next order of the low temperature expansion in $y \gg 1$:
\begin{multline}
 \mathcal F_{\omega_{\mathrm{th}}}(\omega_+)
 =\mathcal F_0
 +\frac{\tan\theta}{\pi}
 \left(\frac{2\pi\omega_{\mathrm{th}}}{\omega_+}\right)
 \\+\frac{\tan\theta}{\tan(\pi\delta)}
 \frac{\delta(1-\delta)}{3(2\delta-1)}
 \left(\frac{2\pi\omega_{\mathrm{th}}}{\omega_+}\right)^2+O\!\left[
 \left(\frac{2\pi\omega_{\mathrm{th}}}{\omega_+}\right)^3
 \right].
 \label{eq:corrected-large-current-fano}
\end{multline}

\begin{figure*}[t]
    \centering
    \includegraphics[width=\textwidth]{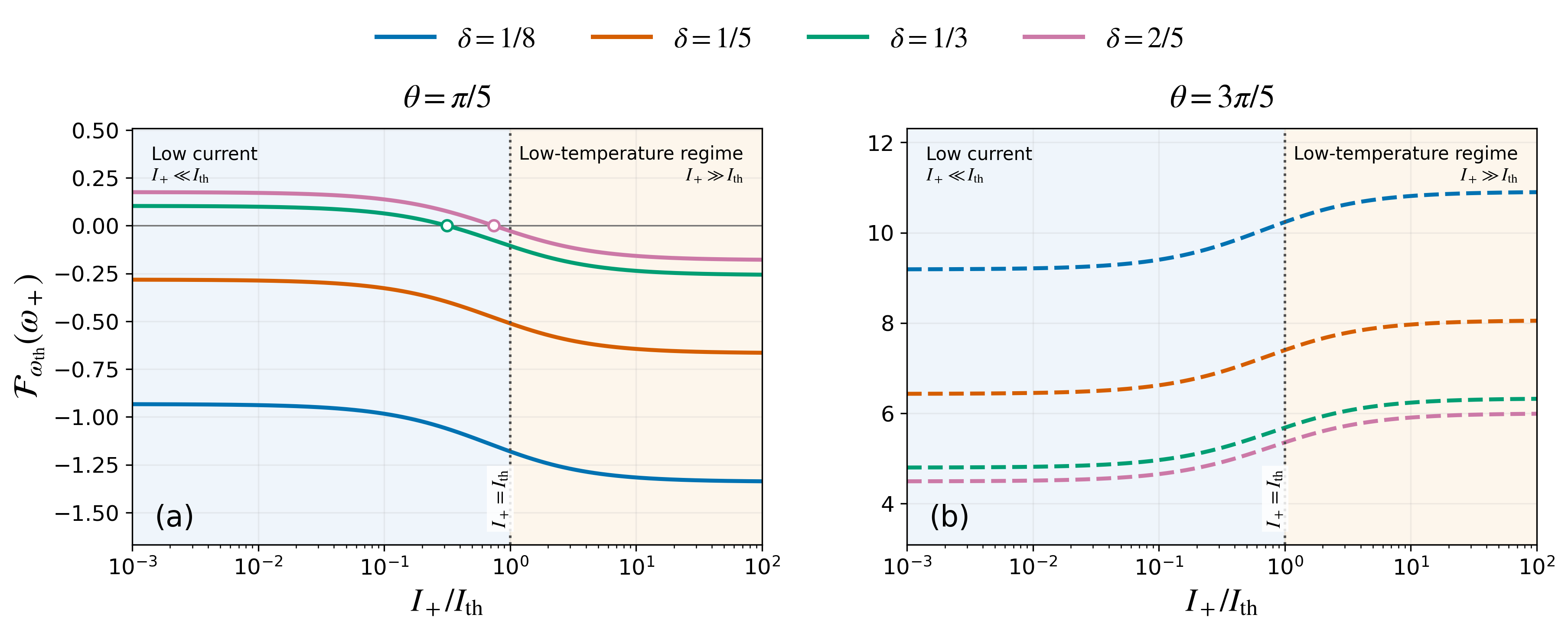}
    \caption{Balanced generalized Fano factor
    $\mathcal{F}_{\omega_{\mathrm{th}}}(\omega_+)$
    as a function of the normalized total injected current
    $I_+/I_{\mathrm{th}}$ at fixed temperature, for
    (a) $\theta=\pi/5$ and (b) $\theta=3\pi/5$.
    Colors indicate the scaling dimensions $\delta$.
    The thermal current scale
    $I_{\mathrm{th}}=\pi e^*\omega_{\mathrm{th}}/\sin^2\theta$,
    with $\omega_{\mathrm{th}}=k_{\mathrm B}T/\hbar$,
    is defined separately for each panel.
    Vertical dotted lines mark $I_+=I_{\mathrm{th}}$,
    separating the low-current ($I_+\ll I_{\mathrm{th}}$)
    and low-temperature ($I_+\gg I_{\mathrm{th}}$) regimes.
    Open circles mark zeros of the Fano factor, occurring
    for $\delta=1/3$ and $2/5$ in panel (a).
    No sign reversal occurs for the curves in panel (b).
    }
    \label{fig:fano-current}
\end{figure*}
Interestingly, the linear term, which originates from the mixed thermal contribution, has a slope $2/(\sin 2\theta )$ with respect to the ratio $e^*\omega_{\rm th}/ I_+$. Remarkably, this coefficient depends only on the statistical phase $\theta$, and not on the scaling dimension $\delta$. Therefore, the leading low-temperature correction to the generalized Fano factor provides, in principle, a direct means of determining $\theta$ independently of $\delta$. It would be interesting to check whether this term could survive beyond the TLL local behavior. 
While present in the Fano factor itself, this linear correction cancels if one considers the extracted local slope in Ref.~\cite{pierre_anyons_PRX_2023}:
\begin{equation}
 P_{\mathrm{slope}}(y)
 =\frac{d[I_+\mathcal F_{\omega_{\mathrm{th}}}(\omega_+)]}{dI_+}
 =\mathcal F_{\omega_{\mathrm{th}}}(y)
 +y\frac{d\mathcal F_{\omega_{\mathrm{th}}}(y)}{dy}.
 \label{eq:corrected-local-slope-definition}
\end{equation}
Consequently,
\begin{multline}
 P_{\mathrm{slope}}(y)
 =\mathcal F_0
 -\frac{\tan\theta}{\tan(\pi\delta)}
 \frac{\delta(1-\delta)}{3(2\delta-1)}
 \left(\frac{2\pi\omega_{\mathrm{th}}}{\omega_+}\right)^2
 \\+O\!\left[
 \left(\frac{2\pi\omega_{\mathrm{th}}}{\omega_+}\right)^3
 \right].
 \label{eq:corrected-large-current-slope}
\end{multline}
But the second-order term provides an additional function of the two unknown parameters that is different from $\mathcal F_0$, thus allowing the determination of both $\theta\mod \pi$ and $\delta$ (with the indeterminacy $\delta\rightarrow 1-\delta$ in case $\theta$ is independent of $\delta$).

%\ipl{We should retain the qualification concerning parameter
%ambiguities here. The full expression~(63) is invariant under
%$\delta\leftrightarrow1-\delta$, so neither its asymptotic
%limits nor its derivatives can distinguish these two values.
%Combining the measurements can therefore constrain $\theta$
%and $\delta$ only up to this symmetry and the phase ambiguity,
%unless additional information fixes the relevant branches.}

Recall that the mixed thermal term $2e^*\omega_{\rm th}G_{\rm coll}(0,\omega_+)$
in the output cross-correlation yields a contribution
$\tan\theta/(\pi y)$ to the balanced Fano factor (see Eq.\eqref{eq:normalized-mixed-thermal-term}). 
At fixed temperature, this contribution is proportional
to $I_+^{-1}$ and therefore cancels in
$P_{\rm slope}=d[I_+\mathcal F]/dI_+$.
Consequently, the leading low-temperature correction to $P_{\rm slope}$ is quadratic in
$\omega_{\rm th}/\omega_+$. 

For $\nu=1/3$ and $\theta=\bar\theta=\pi/3$, one recovers
\begin{equation}
 P_{\mathrm{slope}}
 =-2+\frac{2}{9}
 \left(\frac{2\pi\omega_{\mathrm{th}}}{\omega_+}\right)^2+\cdots,
\end{equation}
whose limiting value agrees with Refs.~\cite{pierre_anyons_PRX_2023,fractional_statistics_gwendal_PRX_2023}.  Since the experimentally
accessible range need not satisfy $y\gg1$, finite-current data should be
compared directly with Eq.~\eqref{eq:corrected-local-slope-definition} using
the full expression \eqref{eq:corrected-finite-T-fano}.

At $\nu=2/5$, one focuses, for a suitable gate-voltage tuning, on the regime where an anyon of charge $e/5$ tunnels between the inner channels\cite{glattli_photo_2018,glattli_imen_2022}
For local tunneling, the conventional
low-energy edge theory gives $\delta=3/5$ and
$\theta=\bar{\theta}=3\pi/5$.
Equation~\eqref{eq:corrected-fano-zero-T} then yields
$\mathcal{F}_0=6$. This positive value does not imply positive output cross-correlations, since the normalization factor $I_+\sin(2\theta)G_{\mathrm{coll}}$ is itself negative for these parameters, with $\sin(2\theta)<0$ and $G_{\mathrm{coll}}>0$. A direct comparison with the experimentally reported Fano factor must therefore account for the different normalization conventions.

As long as $\theta=\pi\delta=3\pi/5$, the finite-temperature Fano factor remains positive and, therefore, cannot account for the experimentally observed negative sign. This discrepancy points to additional effects, such as the dynamics and interactions of the two co-propagating edge modes, edge reconstruction, nonlocal or momentum- and voltage-dependent QPC properties, or the finite width of the injected anyons~\cite{fractional_statistics_width_anyons_rosenow_PRL_2024,martin_anyons_finite_width_FQHE_PRL_2024}. Although the analytic DC solution can be continued to $\delta>1$, this continuation alone does not resolve the sign discrepancy.

An alternative explanation may arise from the diagonal matrix ${\bf K}={\rm diag}(5,5)$, distinct from Wen's construction, which yields properties closely analogous to those at $\nu=1/5$, with a dominant quasiparticle of charge $e/5$ and statistical phase $\theta=\pi/5$ (see Ref.~\cite{ines_statistics_2025}). The physical relevance of this model remains unclear, however, as it requires a small spin polarization or another common quantum number distinguishing the two channels. Nevertheless, irrespective of the interactions between the two modes, $\theta=\pi/5$ yields a negative Fano factor in the low-temperature limit that is close to the value measured in Ref.~\cite{fractional_statistics_gwendal_PRX_2023}. Also, the occurrence of a sign change as $\delta$ crosses the threshold set by Eq.~\eqref{eq:condition_2_5} [see Fig.~\ref{fig:corrected-fano-pi-over-five} for $\theta=\pi/5$] could provide an experimental signature favoring $\theta=\pi/5$ over $3\pi/5$.%\ipl{The phase $\theta=\pi/5$ alone does not guarantee a sign reversal.
%For example, for $\theta=\pi/5$ and $\delta=1/5$, our expression gives
%\[
%\mathcal F_0=-\frac{2}{3},
\qquad
%\mathcal F_\infty\simeq-0.282,
%\]
%and the corresponding curve remains negative throughout the
%crossover, as shown in Fig.~3. Thus, the proposed alternative
%$K$ matrix does not by itself establish a sign reversal:
%we also need to specify the scaling dimension and verify
%condition~(75). The statement ``whatever the interactions are''
%should therefore be qualified.}

\subsection{Low current expansion}
We next examine the domain of sufficiently low injected current at a given temperature, which may be easier to access while maintaining dilute injection:
\begin{equation}
 I_+\ll I_{\mathrm{th}},
 \qquad y\ll1.
\end{equation}
Importantly, this regime does not require increasing the temperature; it can instead be reached by lowering the injected current relative  to the unavoidable finite-temperature scale. Using
\begin{multline}
 \psi(y+\delta)-\psi(y+1-\delta)
 =-\pi\cot(\pi\delta)
\\+y\left[\psi^{(1)}(\delta)
 -\psi^{(1)}(1-\delta)\right]+O(y^2),
 \label{eq:small-y-digamma-expansion}
\end{multline}
we find that the term proportional to $1/y$ generated by the tunneling noise
is exactly canceled by the mixed thermal contribution $2e^*\omega_{\rm{th}}\,G_{\mathrm{coll}}(\omega_-,\omega_+)$ in
Eq.~\eqref{eq:corrected-finite-T-fano}.  The  low current asymptotic behavior is therefore finite:
\begin{equation}
 \mathcal F_{\omega_{\mathrm{th}}}(\omega_+)
 =1-\tan\theta\,R_\delta+O(y),
 \label{eq:corrected-high-T-fano}
\end{equation}
where $R_{\delta}$ is given by :
\begin{equation}\label{eq:Rdelta}
 R_\delta
  =\frac{\tan(\pi\delta)}{\pi^2}
    \bigl[\psi^{(1)}(\delta)
          -\psi^{(1)}(1-\delta)\bigr].
\end{equation}

% beginEq.\eqref{eq:Q-R-definitions}.
%\begin{align}
% C_{\theta,\delta}
% ={}&1-\frac{\tan\theta\tan(\pi\delta)}{\pi^2}
% \left[\psi^{(1)}(\delta)-\psi^{(1)}(1-\delta)\right]
% \nonumber\\
 %={}&1-\tan\theta\,R_\delta.
% \label{eq:corrected-high-T-constant}
%\end{align}
This independence on $I_+$ is due to the fact that both cross-correlations and the denominator in the Fano factor in Eq.\eqref{eq:corrected-balanced-fano-definition} have the same linear behavior with respect to $y$ thus to $I_+$ at $y\ll1$. We thus obtain an additional function of the two unknown parameters, distinct from $\mathcal F_0$, providing a promising route to determine both $\theta$ and $\delta$, up to the ambiguity noted above. For instance, for $\theta=\pi\delta=\pi/3$, the low-current plateaus is given by $-0.14$, thus has a lower amplitude compared to the low-temperature value $-2$. 
The low-current domain must nevertheless be
taken within the window
\begin{equation}
 \omega_+\ll2\pi\omega_{\mathrm{th}}\ll\omega_c,
\end{equation}
where temperature must be high enough to ensure the validity of the perturbative weak backscattering regime whenever $\omega_+$ is made too small.

\subsection{Sign change}

\begin{figure*}[t]
  \centering
  \includegraphics[width=0.95\textwidth]{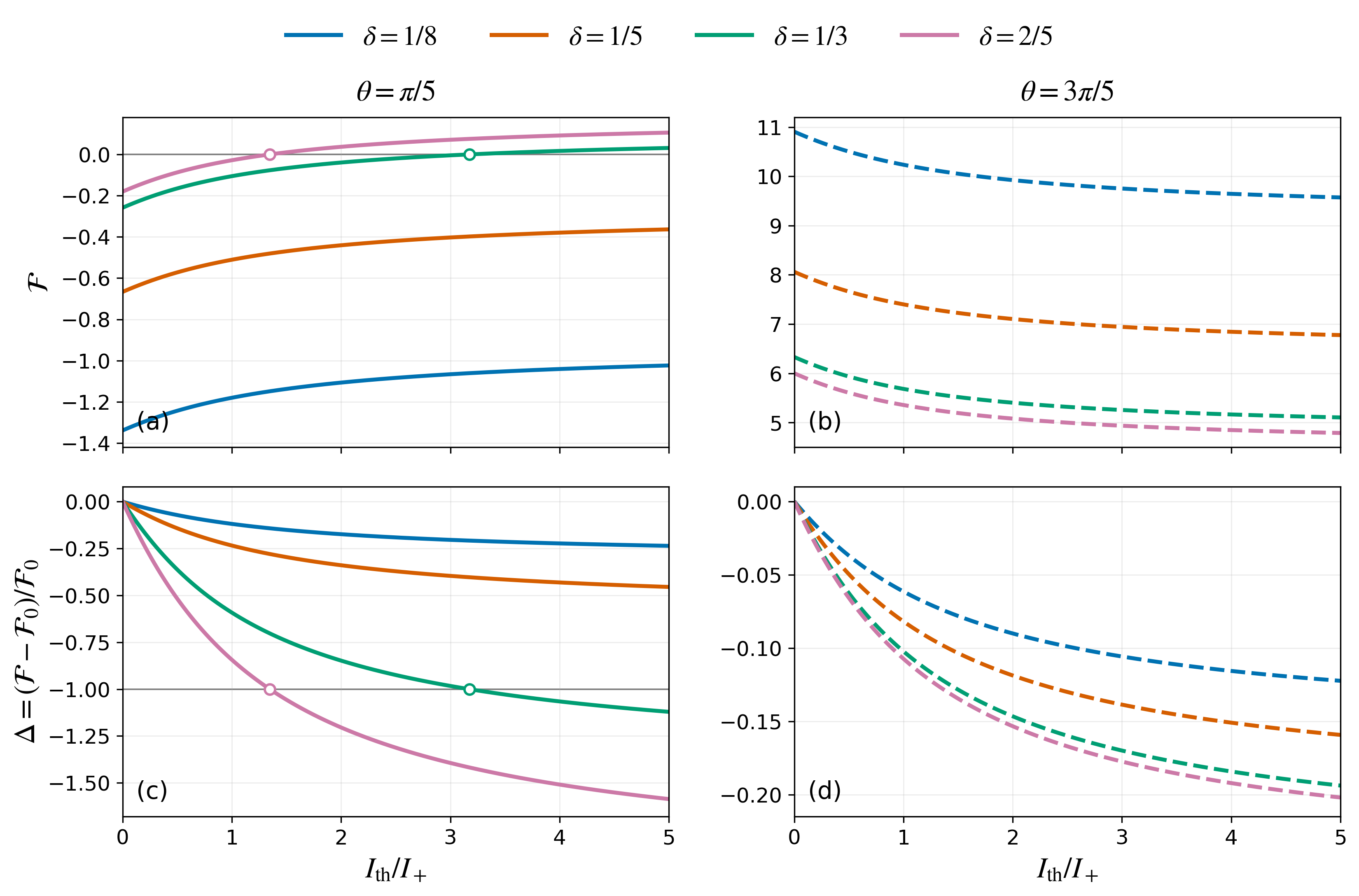}
  \caption{Balanced generalized Fano factor
$\mathcal F_{\omega_{\rm th}}(\omega_+)$ [(a),(b)]
and relative thermal correction
$\Delta_{\omega_{\rm th}}(\omega_+)
=[\mathcal F_{\omega_{\rm th}}(\omega_+)-\mathcal F_0]/
\mathcal F_0$ [(c),(d)] as functions of
$2\pi\omega_{\rm th}/\omega_+=I_{\rm th}/I_+$.
Panels (a),(c) correspond to $\theta=\pi/5$,
and panels (b),(d) to $\theta=3\pi/5$.
Open circles in panels (a),(c) mark the sign reversals
at $I_{\rm th}/I_+\simeq3.174$ for $\delta=1/3$
and $I_{\rm th}/I_+\simeq1.347$ for $\delta=2/5$.
No sign reversal occurs for $\theta=3\pi/5$.
The condition $\Delta_{\omega_{\rm th}}=-1$
is equivalent to $\mathcal F_{\omega_{\rm th}}=0$.}
  \label{fig:corrected-fano-pi-over-five}
\end{figure*}

Figure~\ref{fig:corrected-fano-pi-over-five}
illustrates the thermal crossover of the balanced
generalized Fano factor. At a fixed temperature,
the same crossover can be explored by varying
the injected current $I_+$: decreasing $I_+$
increases the relative temperature $I_{\rm th}/I_+$.

For $\theta=\pi/5$, the curves with $\delta=1/3$
and $\delta=2/5$ exhibit a sign reversal, whereas
those with $\delta=1/8$ and $\delta=1/5$ remain
negative. At fixed $\theta$ and balanced injection,
the normalization $I_+\sin(2\theta)G_{\rm coll}$
has a fixed sign. A zero crossing of the generalized
Fano factor, therefore, also corresponds to a sign
reversal of the output cross-correlation.

For $0<\delta<1$, define
\begin{align}
  Q_\delta
  &=(1-2\delta)\tan(\pi\delta)
  \label{eq:Q-R-definitions}
\end{align}
The values at $\delta=1/2$ are understood in terms of continuity.
Both $Q_\delta$ and  $ R_\delta$ given in Eq.\eqref{eq:Rdelta} are positive and invariant under
$\delta\rightarrow1-\delta$.
The limiting values,
$\mathcal F_0=1-\tan\theta/Q_\delta$ and
$\mathcal F_\infty=1-\tan\theta R_\delta$, 
satisfy $\mathcal F_0<0<\mathcal F_\infty$ when
\begin{equation}
  Q_\delta<\tan\theta<\frac{1}{R_\delta}.
  \label{eq:corrected-sign-reversal-condition}
\end{equation}
This condition ensures at least one zero during
the crossover. For the curves shown in
Fig.~\ref{fig:corrected-fano-pi-over-five},
monotonicity also guaranties that the zero is unique.

For $\tan\theta<0$, the full expression gives
$\mathcal F_{\omega_{\rm th}}(\omega_+)>1$
throughout the crossover, so no sign reversal occurs.
This includes the conventional $\nu=2/5$ assignment
$\theta=3\pi/5$.
For the alternative phase $\theta=\pi/5$ discussed
in Ref.~\cite{ines_statistics_2025}, the left
inequality in
Eq.~\eqref{eq:corrected-sign-reversal-condition}
is satisfied throughout $0<\delta<1$, since
$Q_\delta\leq2/\pi<\tan(\pi/5)$.
The right inequality selects
\begin{equation}\label{eq:condition_2_5}
  \delta_c<\delta<1-\delta_c,
  \qquad
  \delta_c\simeq0.2787,
\end{equation}
where $\tan(\pi/5)R_{\delta_c}=1$.
This interval respects the symmetry
$\delta\rightarrow1-\delta$ and excludes
the locked value $\delta=1/5$.

The reversal results from the competition between
the tunneling noise, the response to the injected
currents, and the mixed thermal contribution;
it does not indicate a change in quasiparticle
statistics. Condition~\eqref{eq:corrected-sign-reversal-condition}
is not satisfied in the locked case
$\theta=\bar\theta=\pi\delta\pmod{\pi}$.
An observed sign reversal would therefore exclude
this locked relation and the $\theta=3\pi/5$
assignment within the assumptions of the present
model, but would not by itself establish an
alternative $K$-matrix description.

\section{Discussion and conclusion}

Let us summarize our analysis. Within a general UNEP framework, we start from the ATE constraint that leads to a fluctuation-dissipation relation, which in turn yields an integral equation relating nonequilibrium DC current and noise~\cite{ines_statistics_2025}. We first enforced the detailed balance equation, without assuming inversion symmetry, so that the DC noise becomes Poissonian. In this regime, the ATE constraint reduces the problem to an integral equation for a single observable-the DC tunneling current-which serves as the distinctive fingerprint of the underlying model.

Solving this equation using the W–H method, we obtained a unique solution: the TLL scaling behavior of the DC current. An alternative method starting from the ATE constraint reinforces this uniqueness (Appendix \ref{ATE+Detailed balance}). Remarkably, although inversion symmetry was not imposed, it is restored by this solution itself. In this sense, TLL scaling is not assumed microscopically but emerges as a necessary consequence of two conditions: the ATE constraint and detailed balance. In turn, when relying on an effective low energy model, the ATE constraint requires spatial locality of the QPC.

We then addressed  the "anyon collider" geometry where the initial density matrix is not thermal. Contrary to previous approaches that assume a TLL description from the outset~\cite{fractional_statistics_theory_2016,fractional_statistics_theory_Sim_Nat_comm_2016,oreg_PRL_FQHE_statistics_2023,fractional_satistics_zhang_gefen_PRL_2025}, we again searched for the class of models compatible with the ATE constraint. In this setup, detailed balance does not hold, and the DC noise becomes super-Poissonian~\cite{ines_cond_mat,ines_PRB_R_noise_2020, fractional_statistics_heiblum_sim_nature_2023}. Therefore, the ATE integral equation involves the two observables, DC current and noise, which do not admit alternative solutions (except TLL) \cite{fractional_statistics_theory_2016,fractional_statistics_theory_Sim_Nat_comm_2016} that satisfy both the ATE constraint and the detailed balance for the reference equilibrium components of the correlators. In addition, we have provided their explicit temperature dependence, showing deviations from the scaling behavior in the two-terminal geometry.
It is important to clarify the physical regimes in which the ATE constraint
can be meaningfully discussed, as it is intertwined with-yet more
robust than-braiding in the time domain. For time-resolved
single-anyon injection, braiding in the time domain with the
statistical phase $\theta$ holds only for copropagating modes. It
further requires either that proper modes have equal velocities, or spatially local injection
at the QPC~\cite{ines_statistics_2025}. In such situations, whenever
braiding holds, the ATE constraint necessarily holds with exactly the same
phase $\bar{\theta}=\theta$. More generally, however, the ATE phase is better protected because it
arises from the exchange of local tunneling operators. 

Cross-correlations in "anyon collider" geometries
\cite{fractional_statistics_gwendal_science_2020,
fractional_statistics_gwendal_PRX_2023,
pierre_anyons_PRX_2023}
have been addressed by relying on a chiral TLL description of the edges, thus implicitly
depend on a scaling dimension $\delta$. In contrast, within the present
framework, the TLL predictions for the DC current and noise emerge as
the unique ones compatible with the ATE constraint. We also require the precise expression of nonequilibrium correlators in terms of their equilibrium counterparts which assumes, as a minimal condition for high order fractional filling factors,  a quadratic low-energy bosonized Hamiltonian. The latter can describe interacting channels where the ATE constraint with the statistical phase $\theta$ is preserved in a local QPC, thus the TLL behavior form of DC transport. It would be interesting to determine whether other models also satisfy the ATE constraint and consequently exhibit the same TLL transport behavior. The edge-TLS toy model in Appendix~\ref{TLS}
illustrates the scope of this conclusion.
For the stationary nonthermal state considered there, coupling to a localized two-level system preserves both ATE and inversion symmetry while altering the tunneling current and noise behavior. This demonstrates that, without detailed balance, these symmetries alone are insufficient to enforce the standard TLL transport behavior.

Experimental deviations from TLL predictions for the DC  current and the Fano factor would therefore signal
that at least one of the underlying assumptions of thermal initial states or the
ATE constraint-is violated. Such deviations could, for instance, arise from
edge reconstruction or from a spatial extension of the QPC. Importantly, a breakdown of the ATE constraint would indicate,
\emph{a fortiori}, a breakdown of braiding in the time domain.

In summary,   when the local exchange relation at the QPC is
combined with detailed balance in a two-terminal setup, or with the
nonequilibrium prescription appropriate to the ``"anyon collider"''
geometry, the resulting expressions for the DC current and noise
reproduce behavior analogous to that obtained within a TLL description
of the edges, once we take the zero-temperature limit. Although these solutions constrain the class of
admissible microscopic models, they do not require the edges to be
described by a free chiral TLL. In particular, hierarchical
co-propagating edge modes interacting on each edge separately lead to the ATE constraint with the robust
statistical phase $\theta$~\cite{ines_statistics_2025}. Its simultaneous
identification with the scaling dimension, established in
\cite{ines_statistics_2025} within a low-energy effective edge model for the edges, has been
generalized in the present work, being based only on the ATE local constraint. This robustness
provides a controlled framework for coherently interpreting future
experimental results and for testing their underlying assumptions or
microscopic models.

An important part of our analysis concerns finite-temperature output cross-correlations in the ``anyon collider'' geometry. Experimentally, the directly accessible observable is not the tunneling noise itself, but the cross-correlated noise between the outgoing edge channels~\cite{ines_bena,ines_gwendal,gu_unpublished}, as demonstrated in the electronic HOM setup~\cite{feve_anyons_martin_Science_2025,ines_alex_2025}. Crucially, our analysis does not rely on the time-domain braiding picture, which involves many non-universal phases when an injected anyon is fractionalized by inter-mode interactions. Instead, it rests on two more robust ingredients: (i) the linear equation of motion governing edge-magnetoplasmon propagation in the interacting channels, which determines the phase-sensitive renormalization of the injected currents, and (ii) the local ATE constraint at the QPC.

These transport behaviors remain compatible with interacting
edge descriptions. In particular, hierarchical
co-propagating modes with short-range interactions
confined to each edge satisfy the local ATE constraint
with $\bar\theta=\theta$~\cite{ines_statistics_2025}.
For the collider geometry, the independent-source
prescription determines the nonequilibrium correlators
from their thermal reference counterparts.

A central application is the finite-temperature output cross-correlation in the "anyon collider", where experiments measure correlations between outgoing channels rather than tunneling noise~\cite{ines_bena,ines_gwendal,gu_unpublished}, as demonstrated in the electronic HOM setup~\cite{feve_anyons_martin_Science_2025,ines_alex_2025}. Our analysis remains valid even for an injected anyon fractionalization by interactions. It relies instead on two robust ingredients: (i) the linear dynamics of edge magnetoplasmons in interacting channels, which determine phase-sensitive renormalization factors for the injected currents, and (ii) the local ATE constraint at the QPC. Thus, the cross-correlations retain their sensitivity to the local exchange phase without requiring a dynamical time-domain braiding interpretation.

On this basis, we extend the ATE solution to the full output cross-correlation and derive the corresponding generalized Fano factor at finite temperature $T=\hbar\omega_{\rm th}/k_B$ and a balanced injection of equal currents $I_+/2$ from the two source QPCs. Given the effective frequency $\omega_+$  $e^*\omega_+=2\sin^2\theta\, I_+$, the resulting Fano factor $\mathcal{F}_{\omega_{\rm th}}(\omega_+)$ depends on the dimensionless ratio  {$\omega_+/\omega_{\rm th}$} as well as on the statistical phase $\theta$ and the scaling dimension $\delta=\bar{\theta}\!\!\!\mod\pi/\pi$. In the zero-temperature limit, it recovers the known Fano factor $\mathcal{F}_0$ of Refs.~\cite{oreg_PRL_FQHE_statistics_2023,fractional_statistics_rosenow_halperin_proceedings_2025}

Finite temperature provides additional information beyond $\mathcal{F}_0$. In the low-temperature regime, the leading correction is $2e^*\omega_{\rm th}/[\sin(2\theta)I_+]$; therefore, it depends only on $\theta$. This first-order correction cancels from the local slope of $I_+\mathcal{F}_{\omega_{\rm th}}(\omega_+)$ with respect to $I_+$, whose leading correction is then of second order and yields an independent function of $\theta$ and $\delta$, complementary to $\mathcal{F}_0$. In the opposite low-injected-current regime, $\omega_+\ll2\pi\omega_{\rm th}$, the Fano factor approaches a plateau that provides a third distinct function of $\theta$ and $\delta$. This latter regime is more readily accessible experimentally than the low-temperature limit, since it can be reached simply by reducing the injected current below an unavoidable finite temperature scale, remaining consistent with dilute, effectively unidirectional injection. The concomitant reduction of the cross-correlation signal may, however, limit experimental precision. Finally, the Fano factor can undergo a temperature-induced sign reversal, originating from the competition between the current-response and collider-noise contributions rather than from any change in quasiparticle statistics.
These results offer additional
ways to constrain $\theta$ and $\delta$, subject
to the phase ambiguity modulo $\pi$ and the
symmetry $\delta\leftrightarrow1-\delta$
at a fixed $\theta$. Measurements of the full finite-temperature crossover,
including a possible sign reversal of the output
cross-correlation, could provide complementary
constraints on quasiparticle statistical phase and edge dynamics.
Combined with independent tests of the ATE relations,
such measurements could help distinguish between
competing descriptions of hierarchical quantum Hall edges.

%%%%%%%% bibliography %%%%%%%%%%%%%
\paragraph{Acknowledgments:} The authors thank Bernd Rosenow, Gwendal Fève, Christian Splatt,  Frédéric Pierre and Anne Anthore for discussions. This work was supported by the ANR grant "QuSig4QuSense" (ANR-21-CE47-0012).
%\bibliography{biblio}
\begin{widetext}
\appendix
\section{UNEP theory conditions.}
\label{app:conditions}
The UNEP theory is valid under three main conditions \cite{ines_cond_mat,ines_PRB_2019}:
\begin{itemize}
    \item (i) the amplitude of $A$ is small, and second-order perturbation theory yields finite results;
    \item (ii) the initial nonequilibrium density matrix ${\rho}_{\mathrm{neq}}$ commutes with $\mathcal{H}_0$, i.e., $[{\rho}_{\mathrm{neq}}, \mathcal{H}_0] = 0$, thus it is stationary, being diagonal with respect to many-body eigenstates of $\mathcal{H}_0$;
    \item  (iii) One has: $\langle A(t)A(0) \rangle = 0$, where $A(t) = e^{\frac{i}{\hbar}\mathcal{H}_0 t} A e^{-\frac{i}{\hbar}\mathcal{H}_0 t}$  in the interaction picture.
\end{itemize}

\section{ATE constraint from the chiral TLL}
\label{app:braiding_TLL}
Here we recall the ATE one gets for a simple fractional filling factor (Laughlin states) for which one has two oppositely chiral edges with bosonic fields labeled by $s=u,d$ (see Fig. \ref{fig.QPC}), obeying\
\begin{eqnarray}
[\phi_{s}(x),\phi_{s'}(x')] = \mp i\pi \delta_{s,s'}\,\text{sign}(x-x'),
\end{eqnarray} 
(where the upper (down) sign is related to $u(d)$ field). The anyon field is given by:
\begin{eqnarray}
\Psi_{s}(x)=e^{\mp i kx+i\sqrt{\nu}\phi_{s}(x)}/\sqrt{2\pi l_{c}}.
\end{eqnarray}
Here, $k$ is the Fermi momentum, and $l_{c} = 2\pi v /\omega_c$ is a short-distance cutoff of the order of the magnetic length, with $v$ is the plasmonic velocity, introducing the frequency cutoff $\omega_c$. Using the Baker-Campbell-Hausdorff formula, one obtains the exchange relation between anyon and anti-anyon fields:
\[
\Psi_{s}(x)\Psi_{s}^{\dagger}(x') = e^{\pm i \pi\nu\,\text{sign}(x-x')} \Psi_{s}^{\dagger}(x')\Psi_{s}(x),
\]
which reveals an exchange fractional phase $\theta=\pi\nu$ in the space domain.  

Let us assume the Hamiltonian ${H}_0$ in Eq.~\eqref{eq:hamiltonian} is formed by two chiral TLLs,   
\begin{eqnarray}
 H_0=\frac{\hbar v}{4\pi}\sum_{s=u,d}\int dx \;(\partial_{x}\phi_{s}(x))^{2},   
\end{eqnarray}
so that one has, in the {interaction} representation, $\phi_{\scriptscriptstyle\mathrm{s,I}}(x,t) = e^{\frac{i}{\hbar}{H}_0 t} \phi_{\scriptscriptstyle\mathrm{s}}(x) e^{-\frac{i}{\hbar}{H}_0 t} = \phi_{s}(x - s vt)$. Choosing spatial locality by setting $x = x'$, one obtains the exchange relation in the time domain:
\begin{eqnarray}\label{eq:braiding_psi}
\Psi_{\scriptscriptstyle\mathrm{s,I}}^{\dagger}(x,t)\Psi_{\scriptscriptstyle\mathrm{s,I}}(x,t') &=& e^{-i\pi \nu \, \text{sign} (t-t')}\nonumber\\
&&\times\,\Psi_{\scriptscriptstyle\mathrm{s,I}}(x,t')\Psi^{\dagger}_{\scriptscriptstyle\mathrm{s,I}}(x,t).
\end{eqnarray}
This shows a similar exchange relation for $A$ in Eq.\eqref{eq:tunA} with a doubled phase: $ A^{\dagger}(t)A(t') = e^{-2i\bar{\theta} \, \text{sign}(t-t')}A(t')A^{\dagger}(t),$ where $\bar{\theta} = \pi \nu$ and we omit the index $\mathrm{I}$ for simplicity.

Notice that this link holds for operators and therefore does not depend on the initial nonequilibrium stationary distribution $\rho_{\mathrm{neq}}$. We then average  this equality with respect to $\rho_{\mathrm{neq}}$, which gives the exchange link in Eq.\eqref{eq:braiding_X}.

\section{Detailed balance equation}
\label{app:detailed_balance}
For a thermal initial distribution, the detailed balance relates the two correlators in Eq.~\eqref{eq:X} as follows:
\begin{multline}
\hbar^{2} X_{\scriptscriptstyle\mathrm{\rightarrow}}(t)
=\langle A^{\dagger}(t)A(0)\rangle
=\frac{1}{Z}\operatorname{Tr}\!\left(
e^{-\beta\mathcal{H}_0}
e^{\frac{i}{\hbar}\mathcal{H}_0t}
A^{\dagger}(0)
e^{-\frac{i}{\hbar}\mathcal{H}_0t}A(0)
\right)
\\
=\frac{1}{Z}\operatorname{Tr}\!\left(
e^{-\beta\mathcal{H}_0}A(0)
e^{\frac{i}{\hbar}\mathcal{H}_0(t+i\hbar\beta)}
A^{\dagger}(0)
e^{-\frac{i}{\hbar}\mathcal{H}_0(t+i\hbar\beta)}
\right)
\\
=\langle A(0)A^{\dagger}(t+i\hbar\beta)\rangle
=\hbar^{2} X_{\scriptscriptstyle\mathrm{\leftarrow}}(t+i\hbar\beta),
\end{multline}
where $Z=\operatorname{Tr}e^{-\beta\mathcal{H}_0}$. Taking the Fourier transformation of this expression yields
\begin{eqnarray}
 X_{\scriptscriptstyle\mathrm{\rightarrow}}(\omega)=\int dt \; e^{i\omega t} X_{\scriptscriptstyle\mathrm{\leftarrow}}(t+i\hbar\beta)=e^{\beta \hbar  \omega}X_{\scriptscriptstyle\mathrm{\leftarrow}}(\omega), \;\; \; \beta=1/\hbar \omega_{\scriptscriptstyle\rm th}.
\end{eqnarray}

\section{Analysis of asymptotics.}
\label{app:asymptotics}
We now analyze the asymptotic behavior of the solution and justify the vanishing of $E(\zeta)$ in Eq.~(\ref{W-H_form}). Since $E(\zeta)$ is analytic in the entire $\zeta$-plane, its possible asymptotic growth at infinity can only be polynomial $E(\zeta) \sim \zeta^{\alpha}$, with an integer exponent $\alpha$. Although the asymptotic form of $F(x)$ is not known \emph{a priori}, the auxiliary functions exhibit well-defined limits at $\zeta \to \infty$, namely 
$|K_{\pm}(\zeta)|\sim \zeta^{\pm(1-2\bar{\theta}/\pi)}$. 
Accordingly, the functions $R_{\pm}(\zeta)$ behave as follows:
\begin{itemize}
    \item For $\bar{\theta}<\pi/2$, the second term dominates, yielding 
    \begin{equation}
    R_{+}(\zeta)\sim
    \frac{i\tan\bar{\theta}}{2}\,
    \pi^{\frac{2\bar{\theta}}{\pi}-1}\,
    \zeta^{-\frac{2\bar{\theta}}{\pi}}, \;\; \;
    R_{-}(\zeta)\sim
    -\frac{i\tan\bar{\theta}}{2}\,
    \pi^{\frac{2\bar{\theta}}{\pi}-1}\,
    \zeta^{-\frac{2\bar{\theta}}{\pi}}.
    \end{equation}
    \item For $\bar{\theta}>\pi/2$, the second term decays faster than $\zeta^{-1}$, giving
    \begin{equation}
    R_{+}(\zeta)\sim
    -\frac{i\tan\bar{\theta}}{2}\,
    \frac{\Gamma(1-\tfrac{\bar{\theta}}{\pi})}{\Gamma(\tfrac{\bar{\theta}}{\pi})}\,
    \zeta^{-1}, \;\;\;
    R_{-}(\zeta)\sim
    \frac{i\tan\bar{\theta}}{2}\,
    \frac{\Gamma(1-\tfrac{\bar{\theta}}{\pi})}{\Gamma(\tfrac{\bar{\theta}}{\pi})}\,
    \zeta^{-1}.
    \end{equation}
\end{itemize}
Hence, for large~$\zeta$, the $\pm$ components of the solution behave as
\begin{equation}
F_{+}(\zeta)=\frac{E(\zeta)+R_{+}(\zeta)}{K_{+}(\zeta)}
\simeq
\frac{E(\zeta)}{K_{+}(\zeta)}
\sim \zeta^{\alpha-1+\frac{2\bar{\theta}}{\pi}},\;\;\; F_{-}(\zeta)=K_{-}(R_{-}(\zeta)-E(\zeta))
\sim \zeta^{\alpha-1+\frac{2\bar{\theta}}{\pi}},
\end{equation}
with $\alpha\in\mathbb{N}_{0}$. Convergence of the
principal-value integral in Eq.~\eqref{eq:equation_current_noise_simple_excess} requires
$\alpha<1-2\bar{\theta}/\pi$ and therefore excludes all
$\alpha\geq1$. For $\bar{\theta}\geq\pi/2$, it also excludes
$\alpha=0$. For $0<\bar{\theta}<\pi/2$, a constant
$E(\zeta)$ would give
$F_\pm(\zeta)\sim\zeta^{2\bar{\theta}/\pi-1}$, which is
incompatible with the normalized Cauchy-projection
asymptotics $F_\pm(\zeta)=O(\zeta^{-1})$. Hence
$E(\zeta)=0$, establishing the uniqueness of the W-H
solution Eq.~(\ref{TLL}). Moreover, any nonzero constant $E(\zeta)=C$ generates an even-in-frequency contribution to the current that dominates the large-frequency asymptotics. Consequently, for one sign of $\omega$, it violates the physical condition $\omega I(\omega)\geq0$, or equivalently makes the noise negative. Spectral positivity therefore requires $C=0$.

\section{Spectral decomposition}
\label{app:spectral}
By using Eq. (\ref{eq:X}) for thermal states, one can obtain 
\begin{align}
\hbar^{2}X_{\scriptscriptstyle\mathrm{\rightarrow}}(\omega)
&=\frac{2\pi}{Z}\sum_{m,n}
|\langle n|A|m\rangle|^{2}
e^{-\beta\hbar\omega_m}
\delta\!\left(\omega-(\omega_n-\omega_m)\right),
\\
\hbar^{2}X_{\scriptscriptstyle\mathrm{\leftarrow}}(\omega)
&=\frac{2\pi}{Z}\sum_{m,n}
|\langle m|A|n\rangle|^{2}
e^{-\beta\hbar\omega_m}
\delta\!\left(\omega+(\omega_n-\omega_m)\right).
\end{align}
where $H_{0}|n\rangle=\hbar \omega_{n}|n\rangle$, from which the detailed balance condition immediately follows. For the zero temperature limit $T \to 0$, the average is taken over the ground state $m=0$. In this case, $X_{\scriptscriptstyle\mathrm{\leftarrow}}(\omega)$ obviously vanishes for $\omega>0$ due to the delta function, and the expression for the tunneling current thus takes the form
\begin{eqnarray}\label{eq:tun_spectral}
 I(\omega_{\scriptscriptstyle\mathrm{dc}})|_{T\to 0}=e^{*}X_{\scriptscriptstyle\mathrm{\rightarrow}}(\omega=\omega_{\scriptscriptstyle\mathrm{dc}})=\frac{2\pi e^{*}}{\hbar^{2}}\sum_{n}|\langle n| A |0\rangle|^{2}\delta\left(\omega_{\scriptscriptstyle\mathrm{dc}}-(\omega_{n}-\omega_{0})\right).    
\end{eqnarray}
On the one hand, this function depends only on $\omega_{\scriptscriptstyle\mathrm{dc}}$. On the other hand, the expression 
Eq.~(\ref{answer}) in the limit $\omega_{\scriptscriptstyle\mathrm{dc}} \gg \omega_{\scriptscriptstyle\rm th}$ yields
\begin{eqnarray}
I(\omega_{\scriptscriptstyle\mathrm{dc}}) \sim \omega_{\rm{th}}G(0)\left(\omega_{\scriptscriptstyle\mathrm{dc}}/\omega_{\scriptscriptstyle\rm th}\right)^{2\delta - 1}.
\end{eqnarray}
The finiteness of this result at $T \to 0$ therefore requires that $G(0)$ exhibits a power-law temperature dependence.

\section{Analysis of solutions for correlation functions}
\label{ATE+DB}
\subsection{ATE and inversion symmetry (without detailed balance)}
\label{ATE+inversion}

Before imposing detailed balance, it is useful to identify what is fixed by the ATE constraint alone. We consider a stationary nonequilibrium state satisfying, in addition to ATE, the inversion symmetry
\begin{equation}
X_{\rightarrow}(t)=X_{\leftarrow}(-t)\equiv X(t),
\end{equation}
or equivalently $X_{\rightarrow}(\omega)=X_{\leftarrow}(-\omega)$. In this case, the DC current is odd in $\omega_{\rm dc}$, whereas the noise is even. The ATE relation reduces to
\begin{equation}
X(t)=e^{-2i\bar\theta {\rm sign}(t)}X(-t).
\label{F1}
\end{equation}
Equation \eqref{F1} constrains only the exchange discontinuity of the correlator under time reversal. It does not determine its complete time dependence. A minimal solution carrying the required monodromy is
\begin{equation}
X_0(t)=C
\left(
\frac{\tau_0}{\tau_0+i t}
\right)^{2\delta},
\qquad
\delta=\frac{\bar\theta}{\pi},
\label{F2}
\end{equation}
where $\tau_0$ is a short time cutoff. Indeed,

\begin{equation}
\frac{X_0(t)}{X_0(-t)}=\left(
\frac{\tau_0+i t}{\tau_0-i t}
\right)^{-2\delta}
\xrightarrow[\tau_0\to0]{}
e^{-2i\pi\delta {\rm sign}(t)}
\end{equation}
which reproduces Eq. \eqref{F1}. The general solution can therefore be written as
\begin{equation}
X(t)=X_0(t)Y(t)
\label{F3}
\end{equation}
where $Y(t)$ is an arbitrary regular function that does not modify
the exchange monodromy. Substituting Eq.~\eqref{F3} into Eq.~\eqref{F1}, one finds that the ATE constraint is preserved provided
\begin{equation}
Y(t)=Y(-t).
\label{F4}
\end{equation}
Thus, ATE, together with inversion symmetry, fixes only the monodromy \eqref{F1}, while leaving an arbitrary even function $Y(t)$. Such factors may encode additional microscopic energy scales or nonequilibrium structure. Consequently, ATE alone does not enforce a pure TLL correlator.

As an explicit example, one can consider
\begin{equation}
Y(t)=e^{-\lambda t^2},
\qquad
\lambda>0.
\label{F6}
\end{equation}
This function is entire and even, and therefore preserves both the ATE phase relation and the analyticity properties of the retarded response function. The corresponding correlator is
\begin{equation}
X(t)=
C
\left(
\frac{\tau_0}{\tau_0+i t}
\right)^{2\delta}
e^{-\lambda t^2}.
\label{F7}
\end{equation}
The associated current and noise read
\begin{equation}
I(\omega)=e^*\bigl[X(\omega)-X(-\omega)\bigr],
\qquad
S(\omega)=e^{*2}\bigl[X(\omega)+X(-\omega)\bigr].
\label{F8}
\end{equation}
The Gaussian factor introduces a new frequency scale $\omega_\lambda\sim\sqrt{\lambda}$. For frequencies $|\omega|\gg \omega_\lambda$, the dominant contribution originates from short times $t\sim |\omega|^{-1}$, where $e^{-\lambda t^2}\simeq1$.
One therefore recovers the usual TLL asymptotics
\begin{equation}
I(\omega)\propto
|\omega|^{2\delta-1}
{\rm sign}(\omega),
\qquad
S(\omega)\propto
|\omega|^{2\delta-1}.
\label{F9}
\end{equation}
In contrast, for $|\omega|\ll\omega_\lambda$, the Gaussian factor suppresses the long-time power-law tail and modifies the low-frequency behavior. Then, \textbf{we obtain} the expression for the current 
\begin{eqnarray}
    I(\omega)=2e^{*}C \tau^{2\delta}_{0} \sin (\pi \delta) \lambda^{\delta-1} \omega \Gamma[1-\delta] {}_1F_1(1-\delta;3/2;-\omega^{2}/4\lambda),
\end{eqnarray}
and for the noise 
\begin{eqnarray}
  S(\omega)=2(e^{*})^{2}C \tau^{2\delta}_{0} \cos(\pi \delta) \lambda^{\delta-1/2}\Gamma[1/2-\delta] {}_1F_1(1/2-\delta;1/2;-\omega^{2}/4\lambda), 
\end{eqnarray}
where ${}_1F_1(a;b;z)$ denotes the confluent hypergeometric function. Using the asymptotic behavior

\begin{equation}
{}_1F_1(a;b;z)
\simeq
\frac{\Gamma(b)}
{\Gamma(b-a)}
(-z)^{-a},
\qquad z\to-\infty,
\end{equation}
one recovers, in the limit $\lambda\to0$, the standard TLL scaling laws.

Therefore, the ATE phase $\bar\theta$ remains unchanged, while the transport properties can be substantially modified. This example demonstrates that ATE combined with inversion symmetry constrains only the exchange monodromy; in the absence of detailed balance, current and noise remain non-universal functions. Within a TLL description of the edges, the ``anyon collider'' provides a physically relevant realization of the general structure in Eq.~\eqref{F3}, as discussed in Ref.~\cite{alex_ines_statistics}.

\subsection{An interacting edge-Two-Level-System (TLS) realization}
\label{TLS}
As an example of a Hamiltonian satisfying both ATE and inversion symmetry, we consider an interacting edge-two-level system (TLS). Localized TLSs have recently been introduced as polarizable fluctuators in compressible strips near quantum Hall edges, where they affect both the edge-plasmon spectrum and the associated heat current~\cite{Sukhorukov_Toma_2025}. Here, we consider instead a simpler, exactly solvable
coupling to the conserved global edge-charge imbalance. Let us denote
\begin{equation}
 A=\xi\psi_u^\dagger(0)\psi_d(0),\qquad
 \hat N_-=\frac{\hat Q_u-\hat Q_d}{2e^*},
 \qquad [\hat N_-,A]=A,
\end{equation}
and consider
\begin{equation}
 H_0=H_{\rm edge}
 +\frac{\hbar\Omega}{2}\sigma_z
 +\hbar g\,\sigma_z\hat N_-,
 \qquad [H_{\rm edge},\hat N_-]=0.
\end{equation}
All expectation values are taken over the factorized
stationary state
\begin{equation}
 \rho=\rho_{\rm edge}\otimes
 \frac{\mathbf{1}_{\rm TLS}}{2},
 \qquad
 \langle\mathcal O\rangle=
 \operatorname{Tr}\!\left(\rho\,\mathcal O\right),
\end{equation}
where
\begin{equation}
 [\rho_{\rm edge},H_{\rm edge}]
 =
 [\rho_{\rm edge},\hat N_-]=0.
\end{equation}
These conditions imply $[\rho,H_0]=0$. This nonthermal factorized state is chosen because a Gibbs state of $H_0$ would generally correlate the TLS with the edge and impose detailed balance. Equal TLS populations yield the even factor $Y_{\rm TLS}(t)$, thereby preserving inversion symmetry. Using $[\hat N_-,A]=A$, one obtains
\begin{equation}
 A(t)=e^{igt\sigma_z}A_0(t),
 \qquad
 A_0(t)=e^{iH_{\rm edge}t/\hbar}
 A e^{-iH_{\rm edge}t/\hbar}.
\end{equation}
The TLS trace then gives
\begin{equation}
 Y_{\rm TLS}(t)
 =
 \frac{1}{2}\operatorname{Tr}_{\rm TLS}
 e^{-igt\sigma_z}
 =
 \cos(gt),
\end{equation}
so that both tunneling correlators acquire the same factor,
\begin{equation}
 X_{\rightarrow}(t)
 =
 X_{\rightarrow}^{(0)}(t)\cos(gt),
 \qquad
 X_{\leftarrow}(t)
 =
 X_{\leftarrow}^{(0)}(t)\cos(gt),
\end{equation}
where $X_{\rightarrow,\leftarrow}^{(0)}$ are evaluated with
$H_{\rm edge}$ and $\rho_{\rm edge}$.

Consequently, if the uncoupled edge correlators obey ATE,
\begin{equation}
 X_{\rightarrow}^{(0)}(t)
 =
 e^{-2i\bar\theta\,\operatorname{sgn}(t)}
 X_{\leftarrow}^{(0)}(t),
\end{equation}
the full correlators obey the same relation. If they also
satisfy inversion symmetry, this symmetry is preserved
because $\cos(gt)$ is even. In frequency space,
\begin{equation}
 X_{\rightarrow}(\omega)
 =
 \frac{1}{2}\left[
 X_{\rightarrow}^{(0)}(\omega+g)
 +
 X_{\rightarrow}^{(0)}(\omega-g)
 \right],
\end{equation}
and similarly for $X_{\leftarrow}$. The resulting
correlator is therefore a superposition of two TLL
contributions shifted by $\pm g$, rather than a
single-scale TLL correlator.

For general diagonal TLS populations, the common factor is
\begin{equation}
 Y_{\rm TLS}(t)
 =
 p_+e^{-igt}+p_-e^{igt}.
\end{equation}
ATE remains valid for arbitrary $p_\pm$, whereas inversion
symmetry requires $p_+=p_-=1/2$. Finally, for $g\neq0$ the
factorized state is generically not a Gibbs state of the
full Hamiltonian, since a global Gibbs state would correlate
the TLS with $\hat N_-$. Thus, no common finite-temperature
detailed-balance condition is imposed in this construction.

\subsection{ATE+Detailed balance}
\label{ATE+Detailed balance}
We are now going to impose the detailed balance relation together with \eqref{eq:braiding_X}. By making the reparametrization in a form
\begin{eqnarray}
X_{\scriptscriptstyle\mathrm{\rightarrow}}(\omega)=e^{+\beta \hbar \omega/2}X(\omega), \quad X_{\scriptscriptstyle\mathrm{\leftarrow}}(\omega)=e^{-\beta \hbar \omega/2}X(\omega).
\label{param}
\end{eqnarray}
Taking the inverse Fourier transform, one can show that the ATE+detailed balance could be reduced to a single equation 
\begin{eqnarray}
X(t+i \hbar \beta/2)=e^{-2i\bar{\theta} \,\text{sign}(t)}X(t-i \hbar \beta/2).  
\label{eq:braiding_X_D_B}
\end{eqnarray}
A minimal solution of Eq.~\eqref{eq:braiding_X_D_B} with the necessary monodromy is obtained by introducing
\begin{eqnarray}
X_0(t)=C_T
\left[
\frac{\pi \tau_0/\hbar\beta}{\cosh(\pi t/\hbar \beta)}
\right]^{2\delta}, \quad \delta=\bar\theta/\pi , \qquad 0<\delta<1 
\end{eqnarray}
Indeed, using
\begin{eqnarray}
\cosh\left(\frac{\pi}{\hbar \beta}\left(t+\frac{i\hbar\beta}{2}\right)\right)
=
i\sinh\left(\frac{\pi t}{\hbar \beta}\right),
\qquad
\cosh\left(\frac{\pi}{\hbar\beta}\left(t-\frac{i\hbar \beta}{2}\right)\right)
=
-i\sinh\left(\frac{\pi t}{\hbar\beta}\right),
\end{eqnarray}
one obtains, with the standard branch prescription,
\begin{eqnarray}
\frac{X_0(t+i\hbar\beta/2)}{X_0(t-i\hbar\beta/2)}
=
\begin{cases}
e^{-2i\pi \delta}, & t>0,\\[2mm]
e^{+2i\pi \delta}, & t<0,
\end{cases}
\label{branch_prescriptions}
\end{eqnarray}
which is precisely Eq.~\eqref{eq:braiding_X_D_B}, since $\bar\theta=\pi \delta$.
More generally, Eq.~\eqref{eq:braiding_X_D_B} allows
\begin{eqnarray}
X(t)=X_0(t)Y(t),
\label{gen_solution}
\end{eqnarray}
where
\begin{eqnarray}
Y(t+i\hbar\beta/2)=Y(t-i\hbar\beta/2),
\qquad \text{or equivalently} \qquad
Y(t+i\hbar\beta)=Y(t).
\end{eqnarray}
Thus, the ATE constraint plus detailed balance fixes the monodromy of the correlator but still allows an additional $i\hbar\beta$-periodic analytic factor. But in contrast to the case of inversion symmetry, 
the imposition of a detailed balance condition fixes $Y(t)=$constant, as will be shown next.

\paragraph{Proposition.}
Let $Y(z)$ be an $i\hbar\beta$-periodic function,
\begin{eqnarray}
Y(z+i\hbar\beta)=Y(z),
\end{eqnarray}
analytic in the strip 
\begin{eqnarray}
\mathcal S=\{z\in\mathbb C:0<\Im(z)<\hbar\beta\},
\end{eqnarray}
and assume that $Y(z)$ has no singularities on the closure of the strip. Suppose furthermore that $Y(z)$ has finite limits at both ends of the strip,
\begin{eqnarray}
\lim_{\Re(z)\to+\infty}Y(z)=Y_+,
\qquad
\lim_{\Re(z)\to-\infty}Y(z)=Y_-,
\end{eqnarray}
uniformly with respect to $\Im(z)\in[0,\hbar\beta]$. Then $Y(z)$ is a constant.

\paragraph{Proof.}
Because of the periodicity $Y(z+i\hbar\beta)=Y(z)$, the upper and lower boundaries of the strip can be identified. Hence $Y$ defines an analytic function on a cylinder. Introduce the variable $z'=\exp\left(2\pi z/\hbar\beta\right)$, then the infinite strip modulo the period $i\hbar\beta$ is mapped to the punctured complex plane $z'\in\mathbb C^\times$. Therefore, there exists a single-valued analytic function $\mathcal{Y}(z')$ such that
\begin{eqnarray}
Y(z)=\mathcal{Y}\!\left(e^{2\pi z/\hbar\beta}\right).
\end{eqnarray}
The limits $\Re(z)\to-\infty$ and \(\Re(z)\to+\infty\) correspond respectively to $z'\to0,
\;
z'\to\infty.$
By assumption, $Y(z)$ has finite limits at both ends of the strip. Hence, $\mathcal{Y}(z')$ has removable singularities at both $z'=0$ and $z'=\infty$. Thus, $\mathcal{Y}(z')$ extends to a holomorphic function on the Riemann sphere. But every holomorphic function on the Riemann sphere is constant. Therefore $\mathcal{Y}(z')=\text{const},$
and consequently $Y(z)=\text{const}.$
\end{widetext}

%\bibliographystyle{apsrev4-2}
%\bibliography{biblio}

\end{document}